\documentclass[aps,
twocolumn,
showpacs,amsmath,amssymb,
superscriptaddress,longbibliography, 10pt]{revtex4-1}
\usepackage{hyperref}

\usepackage{amsmath}
\usepackage{amsfonts}
\usepackage{amssymb}
\usepackage{graphicx}
\graphicspath{{./figures/}}
\usepackage{epsfig}
\usepackage{color}
\usepackage{empheq}
\usepackage{braket}
\usepackage{array}
\usepackage{epstopdf}
\newcommand{\m}{\mathrm}

\newcommand{\eref}[1]{Eq.~(\ref{#1})}
\newcommand{\fref}[1]{Fig.~\ref{#1}}

\definecolor{gold}{RGB}{255,215,0}
\definecolor{blue}{RGB}{0,0,255}
\definecolor{darkgreen}{RGB}{20,150,10}

\begin{document}
\title{Realization of directional amplification in a microwave optomechanical device}

\author{Laure Mercier de L\'epinay}
\email[]{laure.mercierdelepinay@aalto.fi}
\affiliation{Department of Applied Physics, Aalto University, P.O. Box 15100, FI-00076 AALTO, Finland}
\author{Erno Damsk\"agg}
\affiliation{Department of Applied Physics, Aalto University, P.O. Box 15100, FI-00076 AALTO, Finland}
\author{Caspar F. Ockeloen-Korppi}
\affiliation{Department of Applied Physics, Aalto University, P.O. Box 15100, FI-00076 AALTO, Finland}
\author{Mika A.~Sillanp\"a\"a}
\affiliation{Department of Applied Physics, Aalto University, P.O. Box 15100, FI-00076 AALTO, Finland}

\begin{abstract}
Directional transmission or amplification of microwave signals is indispensable in various applications involving sensitive measurements. In this work we show in experiment how to use a generic cavity optomechanical setup to non-reciprocally amplify microwave signals above 3 GHz in one direction by 9 decibels, and simultaneously attenuate the transmission in the opposite direction by 21 decibels. We use a device including two on-chip superconducting resonators and two metallic drumhead mechanical oscillators. Application of four microwave pump tone frequencies allows for designing constructive or destructive interference for a signal tone depending on the propagation direction. The device can also be configured as an isolator with a lossless nonreciprocal transmission and 18 dB of isolation.
\end{abstract} 
\maketitle 


\section{Introduction}

The measurement of weak electromagnetic signals does not only require proper amplification: it is also essential to protect the typically fragile signal source from disturbances by the measurement system. In a typical situation in superconducting quantum information systems working at microwave frequencies, Josephson junction parametric amplifiers are used as a nearly quantum limited readout technology. Unfortunately, these are reciprocal devices, i.e., they amplify signals the same way in either direction, causing the sample to be exposed to a large amplified noise.

One way to break the symmetry between forward and backward transmission is to divide the signal in two branches with transfer phases chosen such that, once recombined, signals propagating in each direction interfere differently. In the microwave domain, this is standardly employed to build isolators and circulators by threading current loops with a magnetic flux. These components are used between the sample and the amplifiers in order to make the signal transmission non-reciprocal. This design however results in bulky devices which are inconvenient in cryogenic systems needed for deep cooling of superconducting quantum systems. Moreover, they employ strong magnetic fields that may perturb sensitive signal sources.

Instead of spatially distinct transfer paths, non-reciprocity can be obtained by a fictitious a loop, which is formed by several modes splitting signals into paths where they interfere under phase-controlled tones driving the system. These ideas have been used in the context of Josephson junction non-reciprocal devices \cite{Abdo2013, Abdo2014, Kamal2014, Sliwa2015, Lecocq2017,Westig2018}, optical nonlinearities \cite{Soljacic2003, Fan2012, Sounas2014, Wu2015, Guo2016, Hua2016} or time-modulation of dielectric constants \cite{Yu2009, Lira2012, Estep2014, Yang2016} where interfering processes generally consist in simultaneous down- and up-conversions.

Based on a slightly different mechanism where counter-propagating optical modes in micro-spheres, -rings or -toroids face different optomechanically-induced transparencies or amplifications, optomechanically-induced nonreciprocity \cite{Hafezi2012} led to a variety of realizations of isolators, circulators and directional amplifiers in the optical domain \cite{Shen2016, Ruesink2016, Shen2018, Ruesink2018}. Even more recently, another route using suitably coupled multimode optomechanical systems \cite{Ranzani2015,Metelmann2015, Fang2017, Li2017,Tian2017,Cheng2018,Li2018} has been investigated. Indeed, the interaction of two optical modes with ancillary mechanical modes also allows to produce multiple interfering transfer paths as required. Notably, this type of system has been adapted recently to promote non-reciprocal coupling between mechanical modes instead and demonstrate a new cooling mechanism \cite{Xu2018}.

Nonreciprocal transfer between two cavities featuring the input and output ports of an optomechanical device can be obtained by balancing their direct coupling with a second transfer path involving one or several mechanical oscillators (MOs), as has been suggested \cite{XuXW2015,Metelmann2015} and experimentally realized \cite{Fang2017}. Ref.~\cite{XuXW2016} proposed a symmetrized scheme where the two paths each incorporate one MO, which was used in recently reported implementations of microwave isolators and circulators  \cite{Peterson2017,Bernier2017, Barzanjeh2017}. While these devices show good isolation, they are not intended to produce gain, and suffer from -- modest -- insertion losses.

Several types of microwave amplifiers based on microwave optomechanical devices have been demonstrated recently \cite{MechAmpPaper,CasparAmp,KippenbergAmp,SqueezeAmp}. The best realizations have achieved a very low noise, even below the quantum limit in a phase-sensitive mode \cite{SqueezeAmp}. All these devices, however, are reciprocal because multimode interfering pathways were not specifically designed. In this work, we demonstrate how one can achieve frequency-converting directional microwave amplifier in a microwave optomechanical system. The system consists of two microwave cavity modes, coupled indirectly via two MOs. Under appropriate driving, each MO creates a frequency-converting amplification path between the cavities as has been previously demonstrated for a single MO \cite{CasparAmp}. Using two MOs, interference between the two paths allows for directional amplification of electromagnetic signals, realizing the scheme proposed in Ref.~\cite{Malz2018}. We further generalize the formalism of Ref.~\cite{Malz2018} to include internal losses of the cavity modes present in the experiment.

\begin{figure}[h]
\includegraphics[width=\columnwidth]{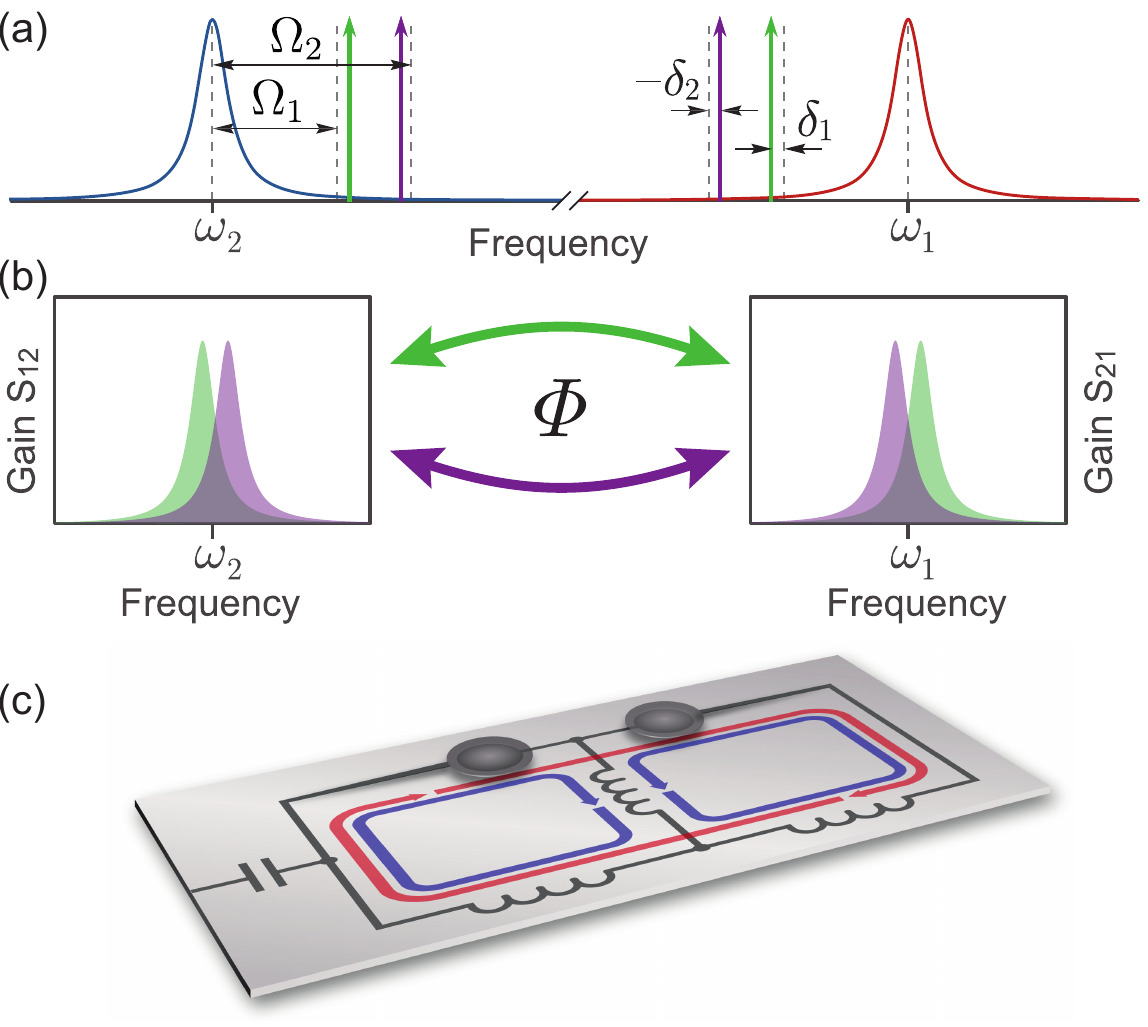}
\caption{\emph{Implementation of directional amplifier.} (a) Four-tone driving scheme. Two optomechanical cavities, which both couple to two mechanical oscillators, are pumped at frequencies close to either red or blue motional sidebands of the mechanical modes. (b) Two-tone driving of each mechanical mode (green or purple tones in (a)) generates bi-directional frequency-converting amplification between cavity modes. Under four-tone driving, the two processes interference is governed by the relative pump phase $\Phi$, enabling directionality. (c) Schematic representation of the device. A superconducting circuit couples two microwave cavity modes (current flow indicated in blue and red) to two mechanical drumhead resonators.}
\label{Fig1}
\end{figure}

\section{Theoretical description}

\subsection{Basic scheme}

The frequency-converting directional amplifier with the pumping scheme schematically illustrated in \fref{Fig1}a  consists of two microwave cavity modes acting as input and output ports, and two MOs mediating two transfer paths for excitations. The mechanical oscillators have the frequencies $\Omega_{1,2}$ and decay rates $\gamma_{1,2}$. The two cavity modes have the frequencies $\omega_{1,2}$, and they couple equally to both mechanical degrees of freedom. The cavities couple to a readout and excitation line with external decay rates $\kappa_{1,2}^e$, and they have the internal decay rates $\kappa_{1,2}^i$. The optomechanical Hamiltonian for the system of two cavities and two MOs is:
\begin{equation}
H_{\rm sys}/\hbar = \sum_i\omega_{i}a_i^\dagger a_i+ \Omega_{i}b_i^\dagger b_i - \sum_{ij} g_{0,ij} a^\dagger_ia_i (b_j+b_j^\dagger)
\label{eq:H}
\end{equation}
where we introduced the intracavity field creation and annihilation operators $(a_i, a_i^\dagger)_{i=1,2}$ and the phononic operators $(b_i, b_i^\dagger)_{i=1,2}$. The parameters $g_{0,ij}$ describe the coupling at the single-quantum level. We suppose that the total linewidths of cavity resonances $\kappa_j = \kappa_j^i+\kappa_j^e$ are much smaller than the mechanical frequencies separation, enabling drive tones to independently address each mechanical mode. Each cavity is coupled to each MO through the excitation of one sideband: red sidebands for the input cavity (hereafter named cavity 1) and blue sidebands for the output cavity (cavity 2), totaling four pump tones as shown in \fref{Fig1}a. The inclusion of blue-sideband drive tones enables amplification, in contrast to previously demonstrated optomechanical isolators \cite{Peterson2017,Bernier2017,Xu2018} which used exclusively drives close to the red sidebands.

The device is pumped at frequencies slightly detuned from cavity 1  red sidebands $\{\omega_1-(\Omega_j-\delta_j)\}_{j=1,2}$ and cavity 2 blue sidebands $\{\omega_2+(\Omega_j-\delta_j)\}_{j=1,2}$
 (see \fref{Fig1}a). The role of the detunings $\delta_{1,2}$ is discussed later. The pumping enhances and linearizes the coupling, with the resulting multiphoton optomechanical coupling energies given by multiplying the single-photon coupling with a given field amplitude. For the red (blue) sideband of MO $j$ on cavity 1 (2), these are denoted as $G_{1j}$ ($J_{2j}$). The cooperativities  corresponding to the red (blue) sidebands $C_{1j}= 4|G_{1j}|^2/(\gamma_{j}\kappa_1)$ (and $C_{2j}= 4|J_{1j}|^2/(\gamma_{j}\kappa_1)$ for blue sidebands) are similar for both MOs $C_{11} \simeq  C_{12}$ (and $ C_{21}  \simeq  C_{22}$ for the blue sidebands). Furthermore, only one phase $\Phi = {\rm Arg}(G_{11})+ {\rm Arg}(J_{21}) - {\rm Arg}(G_{12}) - {\rm Arg}(J_{22})$ is relevant to the amplifier.

\subsection{Model}
We now recall and adapt the formalism developed \cite{Malz2018} in particular to include cavity losses. We show that these losses have some experimentally relevant impact that have to be taken into account in a realistic implementation. The evolution equations for photonic operators involve phononic operators and vice-versa. 
 By eliminating phononic operators (see Appendix A for more details on theory), one is left with coupled photonic equations of evolution, with the particularity that the coupling matrix $T$ is not Hermitian. 
Defining a common vector for intracavity field operators $A\equiv\begin{pmatrix}
a_1&
a_1^\dagger&
a_2&
a_2^\dagger
\end{pmatrix}^{T}$ and corresponding vectors for field input from the external coupler $A_{\rm in}^e$,  noise connected to internal losses $A_{\rm in}^i$ and mechanical noise $B_{\rm in}$ one can define the system susceptibility $\chi[\omega]$ from:
\begin{equation}
A[\omega] = \chi[\omega]\!\cdot\!\left\{ \sqrt{K^e} \!\cdot \!A_{\rm in}^e[\omega] + \sqrt{K^i} \!\cdot \!A_{\rm in}^i[\omega] +  U[\omega] \!\cdot\! B_{\rm in}[\omega]\right\}
\end{equation}
where $\sqrt{K^{e,i}}\equiv{\rm diag}\left(\sqrt{\kappa_1^{e,i}},\sqrt{\kappa_1^{e,i}},\sqrt{\kappa_2^{e,i}},\sqrt{\kappa_2^{e,i}}\right)$ and $U$ is a matrix characterizing  thermomechanical noise impact (see Appendix A). With this definition, the system susceptibility differs from the susceptibility of two bare, uncoupled cavities $\chi_{c,j}[\omega]=\left(\kappa_j/2-i\omega\right)^{-1}$ by the coupling matrix $T[\omega]$:
\begin{equation}
\chi[\omega]^{-1}= {\rm diag}\Big\{
\chi_{c1}^{-1}, \; (\chi_{c1}^*)^{-1}, \chi_{c2}^{-1}, \; (\chi_{c2}^*)^{-1} \Big\}  + T[\omega]
 \label{eq:globalchi}
 \end{equation}
 which is half-empty:
\begin{equation}
T[\omega]= \begin{pmatrix}
T_{11}[\omega] & 0 & 0 & T_{12}[\omega] \\
0 & T_{11}^*[\omega] & T_{12}^*[\omega] &0\\
0 & T_{21}^*[\omega] & T_{22}^*[\omega] & 0\\
T_{21}[\omega] & 0 & 0 & T_{22}[\omega]
 \end{pmatrix},
 \end{equation}
where we used the standard convention $(T_{ij}[-\omega])^*=T_{ij}^*[\omega]$. In the following expressions, we took $C_{11}=C_{12}=C_1$ and $C_{21}=C_{22}=C_2$ though the data is fitted with the general expression allowing these cooperativities to differ slightly. Each $T_{ij}$ element is the sum of two contributions, one from each MO. In the case of $T_{11}$ ($T_{22}$), these contributions represent backactions on cavity 1 (2) from driving both MOs which are added without any multiplying phase factors:
\begin{equation}
\begin{array}{*3{>{\displaystyle}l}}
T_{11}[\omega]&=& \;\;\;C_1\kappa_1/4\left(\gamma_1 \chi_{m1}[\omega] + \gamma_2 \chi_{m2}[\omega]\right)\\[8pt]
T_{22}[\omega]&=& -C_2\kappa_2/4\left(\gamma_1 \chi_{m1}[\omega] + \gamma_2 \chi_{m2}[\omega]\right)\\[8pt]
\end{array}
\end{equation}
with the mechanical susceptibility  $\chi_{m,j}[\omega]=\left(\gamma_{j}/2 - i(\omega+\delta_j)\right)^{-1}$ in the frame rotating at the pump frequencies. In off-diagonal coefficients on the other hand, each of the two contributions accumulates the phase of two different optomechanical interactions. They are then summed with different phases:
\begin{equation}
\begin{array}{*3l}
T_{12}[\omega] \propto \;\;\;\left(e^{i\Phi/2}\gamma_1 \chi_{m1}[\omega] + e^{-i\Phi/2}\gamma_2 \chi_{m2}[\omega]\right)\\[8pt]
T_{21}[\omega] \propto -\left(e^{-i\Phi/2}\gamma_1 \chi_{m1}[\omega] + e^{i\Phi/2}\gamma_2 \chi_{m2}[\omega]\right)
\label{Mij}
\end{array}
\end{equation} 
where the common factor is $\sqrt{C_1C_2\kappa_1\kappa_2}/4$.   
Further defining the output cavity field $A_{\rm out}$ analogously to other vectors and using input-output relations $A_{\rm out}=A_{\rm in}-\sqrt{K^e}\cdot A$, one can get the (optical) transfer matrix $S_{\rm opt}$ defined by $A_{\rm out}=S_{\rm opt} \cdot A_{\rm in}$ when all noise terms are omitted:
\begin{equation}
S_{\rm opt} = \mathbb{I}_4-\sqrt{K^{e}}\cdot \chi[\omega]\cdot \sqrt{K^{e}}.
\label{Smatrix}
\end{equation}
The expression of the non-zero elements of $S_{\rm opt}$ in terms of those of $T$ is cumbersome and can be found in Appendix B. However it is useful to note that $S_{\rm opt}$ has the same zero elements and symmetries as the coupling matrix $T$. For cavity 1, each input operator $a_{\rm in,1}, a_{\rm in,1}^\dagger$ then maps to one of the output operators $a_{\rm out,2}^\dagger, a_{\rm out,2}$ of cavity 2 only, which makes the device a phase-insensitive amplifier \cite{caves82,SqAmpTheory}.
Therefore standard scattering parameters ($S$ matrix) can be defined as transfer amplitudes involving $a_{1, \rm in}$ and $a_{2, \rm out}^\dagger$: $S_{11}=S_{\rm opt}|_{1,1}$, $S_{22}=S_{\rm opt}|_{4,4}$, $S_{12}=S_{\rm opt}|_{1,4}$ and $S_{21}=S_{\rm opt}|_{4,1}$.

\subsection{Working point in lossy cavities }
The off-diagonal elements of $S$ which characterize backward and forward transfer are proportional to the off-diagonal elements of $T$. Isolation ($S_{12}=0$) can therefore be obtained by cancelling $T_{12}$ while keeping $T_{21}$ as large as possible, which will make $T$ and $S_{\rm opt}$ non-Hermitian and $a_1$, $a_1^\dagger$ right eigenvectors of these matrices. 
However, as can be shown from \eref{Mij}, this is only possible at $\omega=0$ if the detunings $\delta_j$ are non-zero: the two effective mechanical susceptibilities' frequency offset arising from pump detunings is then the only source of directionality. Isolation $S_{12}[0]=0$ is furthermore obtained for detunings that compensate the mechanical linewidths imbalance $\delta_{1}=\gamma_1\delta$, $\delta_{2}=-\gamma_2\delta$, and for the phase $\Phi= {\rm Arg}\left( \frac{-1+2i\delta}{\;\,\;1+2i\delta } \right)$. Note that the isolation quality does not depend on cavity losses, and that the interference can appear at a different frequency if the detunings do not exactly compensate for the different mechanical linewidths.

One degree of freedom $\delta$ on the detunings is left:  it is generally \cite{Bernier2017,Peterson2017} tuned to achieve impedance matching of the amplifier to the input line $S_{11}[0]=0$. With non-zero internal cavity losses, this happens when $\delta = \frac{1}{2}\sqrt{C_1\frac{\kappa_1}{\kappa_1^e-\kappa_1^i}-1}$. This can however only be realized for $C_1>0.5\frac{\kappa_1^e-\kappa_1^i}{\kappa_1}$, and if the input cavity is not undercoupled. In our experimental case the lower frequency cavity is undercoupled and is therefore deliberately used  as the output cavity, as seen on \fref{Fig1}a.

When isolation and impedance-matching conditions are satisfied (provided this is possible), the forward power gain of the amplifier is:
\begin{equation}
|S_{21}[0]|^2=\frac{r_2}{r_1}\frac{2C_2(2C_1+1-2r_1)}{(C_1/(2r_1-1)-C_2)^2}
\end{equation}  
where we introduced $r_j=\kappa_j^e/\kappa_j$. 
As previously observed for the single-MO amplifier, the cooperativities are best chosen both large and nearly equal while maintaining $C_2<C_1+1$ to prevent mechanical instability. This expression coincides with the gain calculated in \cite{Malz2018} in the limit of non-lossy cavities. However $C_1$ now compares to a reduced cooperativity $(2r_1-1)C_2$ in the denominator so that only a finite gain can be obtained at the onset of instability, contrary to the single-MO amplifier \cite{CasparAmp}.
The gain at instability, which is the maximum gain achievable with these constraints, is $\sqrt{\frac{r_2}{r_1}}\frac{2r_1-1}{1-r_1}$ in the limit of large cooperativities. Output cavity losses (low $r_2$) are less detrimental to the gain than input cavity losses, which is a second reason for using our undercoupled cavity on the output side. 
A more detailed analysis of the impact of cavity losses on gain is given in Appendix B.
While the impedance matching condition is required to realize an ideal quantum-limited amplifier, it restricts the choice of operating parameters and in a practical device a better trade-off may be possible. For example, for our experimental parameters, the maximum gain while strictly enforcing impedance matching and perfect isolation conditions would be $-2.6$ dB. By relaxing the impedance matching condition we can nevertheless realize directional amplification. We note that impedance mismatch also reduces backward-propagating added noise of the amplifier  \cite{Malz2018}. 

\section{Experimental implementation}

\subsection{Experimental details}
Our device is fabricated by patterning a microwave circuit and mechanical oscillators in aluminum on a quartz substrate. The mechanical elements are circular membranes evaporated on top of a sacrificial amorphous silicon layer, which is then removed using isotropic reactive plasma etching to release the membranes. The two drumheads of diameters $19.7$ and  $16.9 \rm \mu m$  vibrate above circular electrodes to form displacement-dependent capacitors \cite{Teufel2011} with fundamental frequencies $\Omega_{1}/2\pi=9.24\,\rm MHz$ and $\Omega_{2}/2\pi=9.82\,\rm MHz$ and decay rates $\gamma_1/2\pi \simeq 310\,\rm Hz$ and $\gamma_2/2\pi \simeq 290\,\rm Hz$. The microwave circuit sustains two eigenmodes with frequencies $\omega_1/ 2\pi = 3.89\,\rm GHz$ and $\omega_2/ 2\pi =5.63\,\rm GHz$ that couple roughly equally to both mechanical degrees of freedom, see \fref{Fig1}c. The two cavity modes couple to a line used for both and readout and excitation with external decay rates $\kappa_1^e/2\pi=  406\,\rm kHz$ and $\kappa_2^e/2\pi = 115\,\rm kHz$ (see \fref{Fig1}), and internal decay of $\kappa_1^i/2\pi = 197\,\rm kHz$ and $\kappa_2^i/2\pi=  233\,\rm kHz$ respectively, which makes cavity 1 overcoupled and cavity 2 undercoupled to the feedline. As supposed in the modeling, the total cavity linewidths $\kappa_j = \kappa_j^i+\kappa_j^e$ are smaller than the mechanical frequencies separation, enabling the drive tones to drive a single process at a time.

\begin{figure}[h]
\includegraphics[width=\columnwidth]{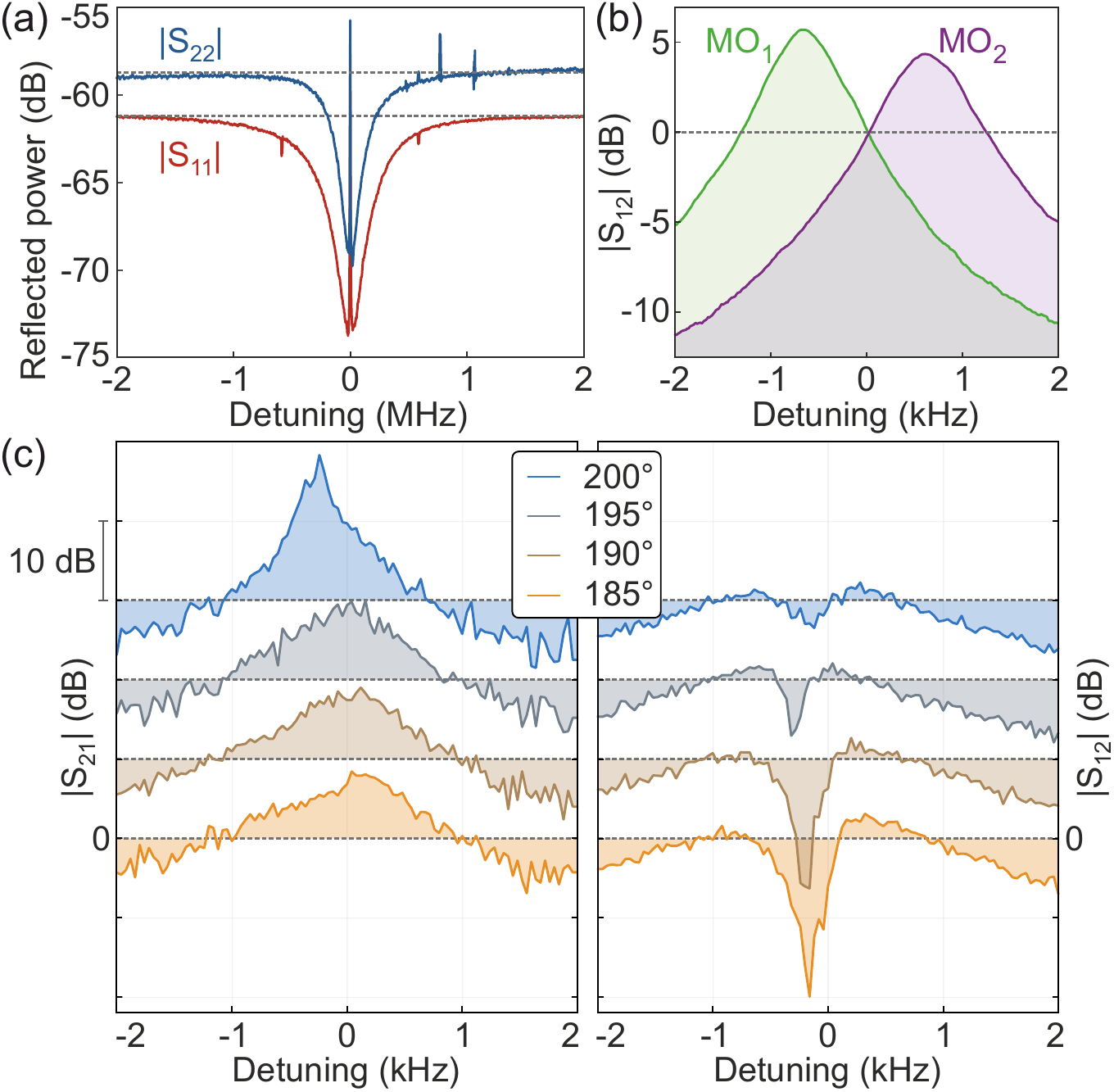}
\caption{\emph{Preparation of interference of amplification channels.}  (a) Power reflection on cavity 1 (red) and 2 (blue)  over a broad frequency span around  the resonance frequency of either cavity. Here, two independent single-MO amplification processes are active (see text). Cavity response is used to determine pump and readout efficiencies through the whole cryogenic attenuation and amplification stage around each frequency range, indicated as dashed lines: $-61.2\,\rm dB$ around cavity 1, $-58.8\,\rm dB$ around cavity 2. Small features detuned by about $0.6\,\rm MHz$ from cavities frequencies correspond to field oscillations at the mechanical frequency difference. Larger spurious peaks on $|S_{22}|$ above the cavity frequency arise from intermodulation of the pump tones in the measurement system. (b) Backward power transmission $|S_{12}|$ of single-MO amplifiers built from MO1 only (green) and MO2 only (purple), once balanced, with respect to frequency detuning from cavity 1 (the output frequency in this case). (c) Forward $|S_{21}|$ and backward $|S_{12}|$ transmission for different phase $\Phi$ values of one of the generators relative to the others.}
\label{Fig2}
\end{figure}
The system is operated in a dilution refrigerator at a fixed temperature of $200\,\rm mK$. The elevated temperature was chosen because we found that the mechanical frequencies were fluctuating at the base temperature, and an accurate drive tone detuning could not be maintained. Four synchronized independent generators, whose relative phase drift is smaller than $4^\circ$ per hour, are used to pump the device.

We record the transfer parameters using a Rohde \& Schwartz ZVA-50 network analyzer that allows independent excitation and measurement frequencies. The probe is maintained at a very low power, $-76\,\rm dB$ below the lowest pump power to ensure that probing does not modify the amplifier's behavior.
An independently measured contribution by noise in the recorded frequency-converting response is subtracted to yield the pure transfer coefficients as explained in Appendix D.
The pump and probe efficiencies around the cavity frequencies are determined by measuring a large frequency span around the cavities (see \fref{Fig2}a) and used to calibrate the transfer parameters of the amplifier following the method discussed Appendix E.

\subsection{Directional amplifier}

To prepare the interference effect,  we first drive each MO independently through its red and blue sideband, reproducing two single-MO optomechanical reciprocal amplifiers as described in \cite{CasparAmp}. The blue sideband drive powers are tuned to produce similar amplification for both single-MO amplifiers in the backward direction $S_{12}$, as shown on \fref{Fig2}b. In the experimental situation, the cooperativities employed are different for each MO, $C_{11}=1.27$, $C_{12}=3.20$ for the red sidebands  of MO1 and 2 respectively, and $C_{21}=1.33$, $C_{22}=2.05$ for the blue sidebands. The pump tones are detuned by $\delta_1/2\pi=-\delta_2/2\pi=600\,\rm Hz$ from the sidebands. We then turn on all pumps simultaneously  and the phase of one of them relative to the others is tuned to achieve destructive interference in the backward direction as shown in \fref{Fig2}c. Further fine tuning of the frequencies, phase and powers is generally required to compensate for slow phase and power drift of the generators and slight power dependence of the cavity frequencies.

\begin{figure}[h]
\centering
\includegraphics[width=\columnwidth]{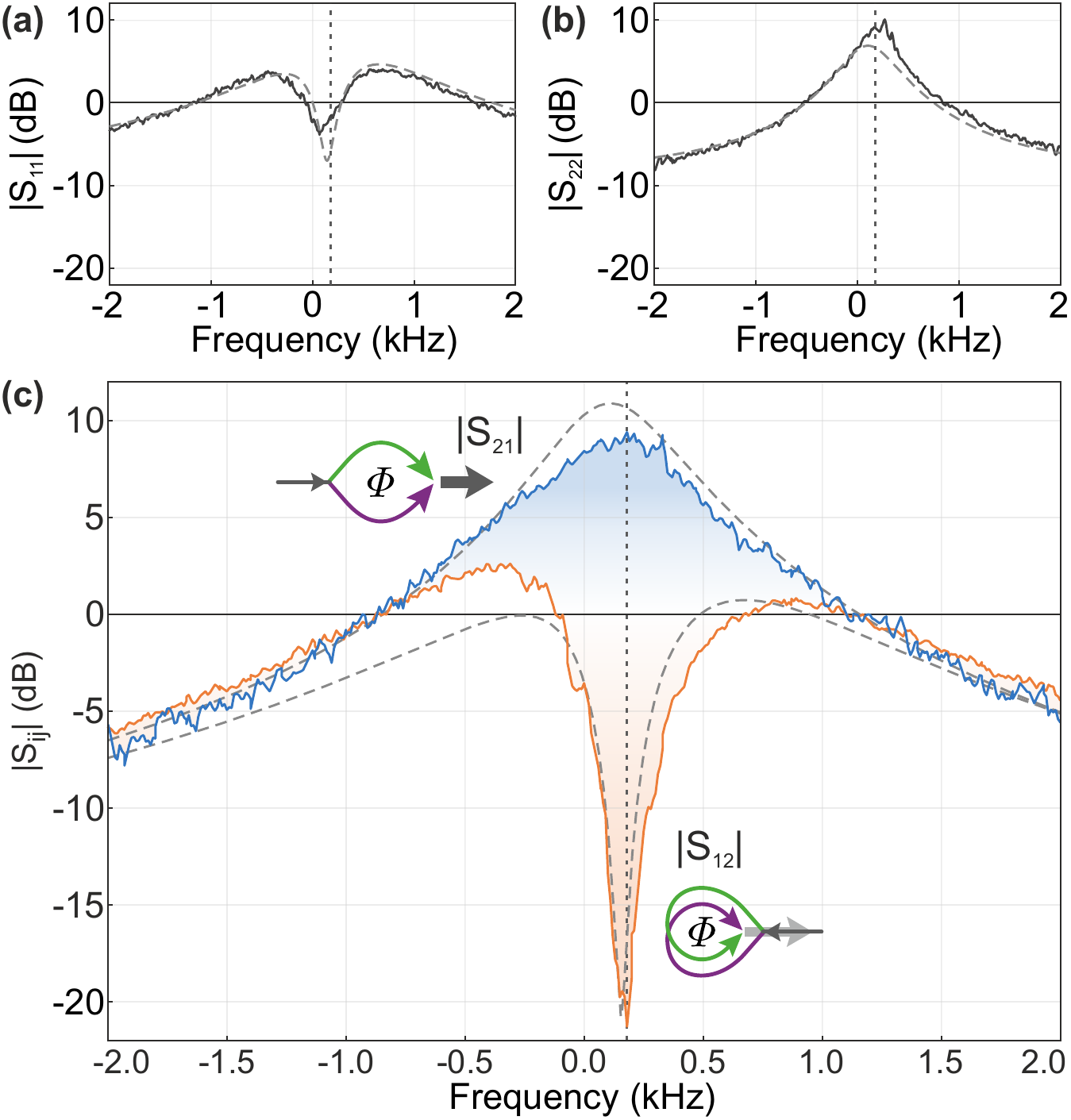}
\caption{\emph{$S$-parameter amplitudes of the amplifier with respect to frequency.} (a) $|S_{11}|$, and (b) $|S_{22}|$, (c) $|S_{12}|$ (orange) and $|S_{21}|$ (blue). Grey dashed lines are simultaneous fits of all 4 data sets with the expressions of the text, where the phase is left free. The amplifier bijectively maps a frequency range around one cavity resonance to a mirrored range around the other cavity resonance: only one increasing frequency range around $\omega_1/2\pi$ is used as the horizontal axis of all plots for comparison. The maximum non-reciprocity frequency is indicated with a dashed vertical line. }
\label{Fig3}
\end{figure}

The amplifier maps bijectively a frequency range around $\omega_1/2\pi$ to a  frequency range centered and mirrored about $\omega_2/2\pi$. A minimum backward transmission gain of $-21.3\pm1.1\,\rm dB$ is observed, as displayed in \fref{Fig3}, while the forward gain reaches $9.4\pm1.1\,\rm dB$  at the same frequency. The uncertainty corresponds to the estimated maximum gain calibration uncertainty (see Appendix E). The maximal non-reciprocity factor $|S_{21}/S_{12}|$ is therefore $30.7\pm2.2\,\rm dB$ which compares well to isolator implementations \cite{Peterson2017, Bernier2017}. While perfect impedance matching was not enforced, the reflection attenuation ($S_{11}$) reaches $3.9\,\rm dB$. The isolation bandwidth, defined as the frequency range where the backward transmitted power is attenuated by more than $3\,\rm dB$, is found be $500\,\rm Hz$, as expected since it is governed by the mechanical linewidths. Taking the amplification bandwidth as the frequency range amplified by more than half the maximum gain, the latter amounts to $675 \,\rm Hz$, also of the order of magnitude of mechanical linewidth because of the interference effect on the forward transfer. 

\subsection{Lossless isolator}

\begin{figure}
\centering
\includegraphics[width=\columnwidth]{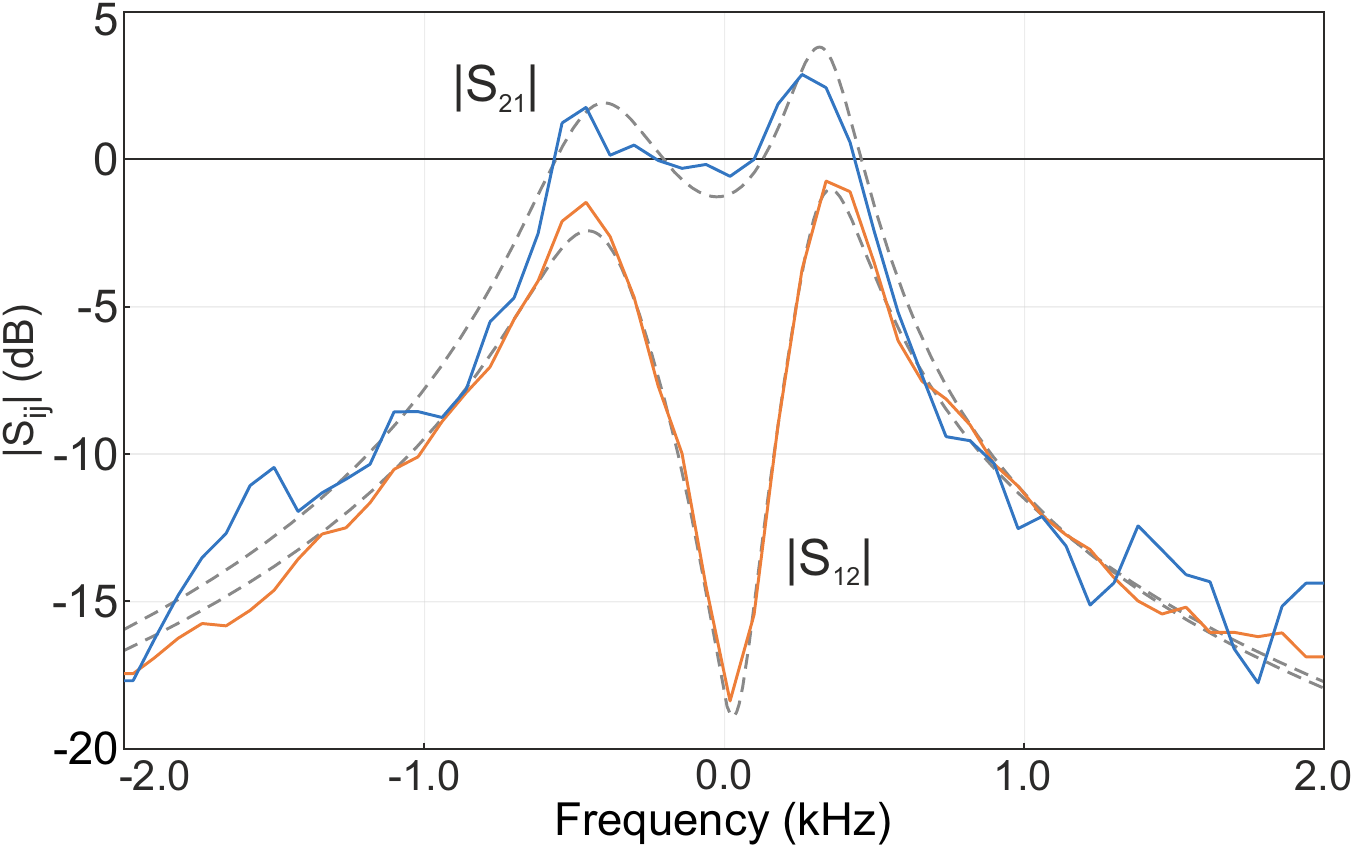}
\caption{$|S_{12}|$ (orange) and $|S_{21}|$ (blue) transfer parameters of a lossless isolator. Grey dashed lines are fits with the expressions of the text, where the phase is left free.}
\label{Fig4}
\end{figure}

In another configuration ($C_{11}=0.47$, $C_{12}=0.74$, $C_{21}=0.84$, $C_{22}=0.96$, $\delta_1/2\pi= 450\,\rm Hz$, $\delta_2/2\pi=- 405\,\rm Hz$), the same device can be used as an isolator free from any insertion loss whose forward and backward transfer gains are represented on \fref{Fig4}. In this configuration, we obtain an isolation by $18.0\,\rm dB$ of backward-propagating signals.

\section{Discussion} 

Generating interference between cavities coupling paths requires mechanical oscillators with similar linewidths and optomechanical couplings.
However, non-reciprocal multimode devices also require very different mechanical frequencies -- in the sense that the spacing of the frequencies generally needs to be a rather large fraction of their value --  to allow for each mechanical oscillator to be addressed separately. This condition, that one could call RSBD (Resolved SideBands Difference) analogously to RSB (Resolved SideBand) appears to be very relevant as parasitic cross-driving of mechanical modes has been shown to significantly increase optomechanical isolators insertion loss \cite{Bernier2017}. The present scheme balances these opposite requirements of equal couplings but different frequencies by involving two separate drum resonators, contrary to what has been done previously in \cite{Peterson2017,Bernier2017, Barzanjeh2017} where different eigenmodes of the same resonator were used. For this reason this scheme is also immune to any direct coupling between modes through geometrical nonlinearities.

While the current device represents a technological step forward in multimode optomechanical applications, the RSBD condition was attained here at the price of deliberately reducing the cavities external decay rates, thus departing from the ideal far-overcoupled situation. The question of cavity dissipation might in fact become essential in the description of real multimode optomechanical systems, which is why the present article attempts to draw a particular attention to them. On the experimental side, achieving the RSBD condition while maintaining overcoupled cavities is one of the next endeavors to manufacture high quality non-reciprocal devices from multimode optomechanical systems. Another future goal is to carefully characterize the noise properties of the amplifier. One can realistically reach an added noise near the quantum limit, however, this requires operation at temperatures appreciably lower than what we used in the current experiment for stability reasons.

\section{Conclusions}

We have reported on a multimode optomechanical directional amplifier, both demonstrating a high isolation between forward and backward transfer and gain in the forward direction. The work also demonstrates a non-reciprocal optomechanical device using two separate mechanical resonators. Moreover, this work allowed to formulate some guidelines for the design of such devices in the real situation where cavities dissipation rates compare to other system frequencies.

\begin{acknowledgments} This work was supported by the Academy of Finland (contracts 250280, 308290, 307757, 312296), by the European Research Council (615755-CAVITYQPD), and by the Aalto Centre for Quantum Engineering. We acknowledge funding from the European Union's Horizon 2020 research and innovation program under grant agreement No.~732894 (FETPRO HOT). We acknowledge the facilities and technical support of Otaniemi research infrastructure for Micro and Nanotechnologies (OtaNano).
\end{acknowledgments}

\section*{Appendices}

\appendix

\section{Theory}

\subsection{General expression for the coupling matrices}
Creation and annihilation operators for cavity $j$ are denoted $a_j$ and $a_j^\dagger$, same operators for mechanical oscillator (MO) $j$ are denoted $b_j$ and $b_j^\dagger$. The total Hamiltonian of the system is given in \eref{eq:H}.

Following \cite{Malz2018}, we consider for generality a more complete pumping scheme than what is used in the article. In the complete scheme, each cavity $i$ can be driven by 4 tones with frequencies $\{\omega_{i} - (\Omega_j+ \delta_j)\}_{j=1,2}$, that is, the frequencies of the two sidebands of each of the two oscillators. The photon field $a_i(t)$ in cavity $i$ can be decomposed into  a coherent driven part of amplitude $\alpha_{i}(t)$ oscillating at pump frequencies and a fluctuation $\delta a_i(t)$:
%
\begin{equation}
\begin{array}{*3{>{\displaystyle}l}}
a_i(t)&=& e^{-i\omega_i t}\alpha_i(t) +\delta a_i(t) \\[6pt]
\alpha_{i}(t)&\equiv& \sum_j \alpha_{ij-}\,e^{-i(\Omega_j-\delta_j)t} + \alpha_{ij+}\,e^{-i(\Omega_j+\delta_j)t},
\end{array}
\end{equation}
where we introduced the intracavity field amplitudes at the frequencies of the red-detuned (blue-detuned) sidebands $\alpha_{ij-}$ ($\alpha_{ij+}$).
The Hamiltonian can be linearized in the fluctuations $\delta a_i$ in a standard semi-classical approach. The quantum photon field fluctuation $\delta a_i$ is now renamed $a_i$ for convenience.
In the frame rotating with:
\begin{equation}
H_{0}/\hbar = \sum_i\omega_{i}a_i^\dagger a_i+ \sum_j(\Omega_{j}+ \delta_j) b_j^\dagger b_j,
\end{equation}
the linearized Hamiltonian is rewritten:
\begin{multline}
H_{\rm rot}/\hbar=-\sum_{j}\delta_jb_j^\dagger b_j - \sum_{ij}g_{0,ij}\Big\{\alpha_{i}^*(t)a_i\times \\ \left[b_je^{-i(\Omega_j+\delta_j)t}+b_j^\dagger e^{+i(\Omega_j+\delta_j)t}\right]
+H.c\Big\}
\label{eq:beforeRWA}
\end{multline}
The hierarchy of the system frequencies is: $\delta_{1,2}\simeq\gamma_{1,2}\ll\kappa_{1,2}<|\Omega_2-\Omega_1|<\Omega_{1,2}\ll\omega_{1,2}$. The system is in the resolved-sideband regime. However some terms in the last equation are also oscillating at the mechanical frequency difference that is only greater than the cavities linewidths by some tens of percent. \\

Neglecting any term oscillating faster than $\gamma_j$ in the evolution equation for phononic operators $b_j$ and any term oscillating faster than $\kappa_i$ in the evolution equation for photonic operators $a_i$, the Fourier transform of the evolution equations gives: 
\begin{equation}
\left\{\begin{array}{*3{>{\displaystyle}l}}
b_j[\omega]&=&\chi_{m,j}[\omega] \Big(i\sum_{i=1}^2g_{0,ij}\Big\{\alpha_{ij-}^*a_i[\omega]+\alpha_{ij+}a_i^\dagger[\omega]\Big\} \\[4pt]
 & &\hfill +\sqrt{\gamma_j}\,b_{{\rm in},j}[\omega]\Big)  \\
a_i[\omega]&=& \chi_{c, i}[\omega] \Big( i\sum_{j=1}^2g_{0,ij}\Big\{ \alpha_{ij+} b_j^\dagger[\omega] + \alpha_{ij-}b_j[\omega] \Big\}\\[4pt]
& & \hfill +\sqrt{\kappa_{i}^{i}}a_{{\rm in},i}^{i}[\omega]+\sqrt{\kappa_{i}^{e}}a_{{\rm in},i}^{e}[\omega]\Big)\label{eq:aomega}\\[10pt]
\end{array}\right.
\end{equation}
with the effective mechanical and cavity susceptibilities:
\begin{equation}
\left\{\begin{array}{*3{>{\displaystyle}l}}
\chi_{m,j}[\omega]&\equiv&\left(\frac{\gamma_j}{2}-i(\omega+\delta_j)\right)^{-1}\\[10pt]
\chi_{c, i}[\omega]&\equiv&\left(\frac{\kappa_{i}^{i}+\kappa_{i}^{e}}{2}-i\omega\right)^{-1}.\\
\end{array}\right.
\end{equation}
Here we used the same convention for operators and functions: for any annihilation operator $(c[\omega])^\dagger=c^\dagger[-\omega]$ (and for any function $(f[\omega])^*=f^*[-\omega]$). 

The effective couplings to each cavity that appear in the previous equations are now written $G_{ij}\equiv g_{0,ij}\alpha_{ij-}$ and $J_{ij}\equiv g_{0,ij}\alpha_{ij+}$.  $G_{ij}$ ($J_{ij}$) therefore corresponds to the enhanced optomechanical coupling from red-sideband (blue-sideband) pumping, the first index is attached to the cavity and the second to the MO. \\

We now simplify the problem involving 4 photonic and 4 phononic operators to a problem involving only 1 photonic and 1 phononic vector operators. To this end we define $A[\omega]\equiv\begin{pmatrix}
a_1[\omega]&
a_1^\dagger[\omega]&
a_2[\omega]&
a_2^\dagger[\omega]
\end{pmatrix}^{T}$ and  $B\equiv\begin{pmatrix}
b_1[\omega]&
b_1^\dagger[\omega]&
b_2[\omega]&
b_2^\dagger[\omega]
\end{pmatrix}^{T}$ and corresponding $A_{\rm in}^e[\omega]$, $A_{\rm in}^i[\omega]$ and $B_{\rm in}[\omega]$. We also define the susceptibility matrices for the cavities and mechanical oscillators:
\begin{equation}
\left\{\begin{array}{*3{>{\displaystyle}l}}
\Xi_c[\omega] &\equiv& {\rm diag}\Big\{
\chi_{c,1}[\omega], \chi_{c,1}^*[\omega], \chi_{c,2}[\omega], \chi_{c,2}^*[\omega] \Big\} \\[5pt]
\Xi_m[\omega] &\equiv& {\rm diag}\Big\{
\chi_{m,1}[\omega], \chi_{m,1}^*[\omega], \chi_{m,2}[\omega], \chi_{m,2}^*[\omega] \Big\},\\[5pt]
\end{array}\right.
\end{equation}
the total cavity and mechanical decay rates vectors:
\begin{equation}
K \equiv {\rm diag}\Big\{
\kappa_{1}, \kappa_{1}, \kappa_{2}, \kappa_{2} \Big\}\qquad \Gamma \equiv {\rm diag}\Big\{
\gamma_{1}, \gamma_{1}, \gamma_{2}, \gamma_{2} \Big\}
\end{equation}
and $K^{i}$ and $K^{e}$ the internal and external cavity decay rates matrices with analogous definitions.
Having performed this matrix formalization, we can now rewrite the two coupled equations for photonic and phononic fields:
\begin{equation}
\begin{array}{*3{>{\displaystyle}l}}
A[\omega]&=& \Xi_c[\omega]\cdot \Big(  \mathcal{G}  \cdot B[\omega] + \sqrt{K^e}  \cdot A_{\rm in}^e[\omega]+ \sqrt{K^i } \cdot A_{\rm in}^i[\omega] \Big)\\
B[\omega]&=&  \Xi_m[\omega]\cdot \Big(  \mathcal{H}  \cdot A[\omega] + \sqrt{\Gamma}  \cdot B_{\rm in}[\omega] \Big)
\end{array}
\end{equation}
where the coupling matrices are: 
\begin{equation}
\mathcal{G} \equiv i\begin{pmatrix} 
G_{11} & J_{11} & G_{12}  & J_{12}\\
-J_{11}^* & -G_{11}^* & -J_{12}^* &- G_{12}^*\\
G_{21} & J_{21} & G_{22}  & J_{22}\\
-J_{21}^* & -G_{21}^* & -J_{22}^* &- G_{22}^*
\end{pmatrix},
\end{equation}
\begin{equation}
\mathcal{H} \equiv i\begin{pmatrix} 
G_{11}^* & J_{11} & G_{21}^*  & J_{21}\\
-J_{11}^* & -G_{11} & -J_{21}^* &- G_{21}\\
G_{12}^* & J_{12} & G_{22}^*  & J_{22}\\
-J_{12}^* & -G_{12} & -J_{22}^* &- G_{22}
\end{pmatrix}.
\end{equation}
Replacing the phononic matrix operator in the photonic matrix equation, one gets the expression of photonic fields perturbated by the coupling to phononic fields:
\begin{multline}
\Big( \Xi_c[\omega]^{-1} - \mathcal{G}  \cdot \Xi_m[\omega] \cdot  \mathcal{H} \Big) A[\omega]=   \\  \sqrt{K^e}  \cdot A_{\rm in}^e[\omega]+ \sqrt{K^i }  \cdot A_{\rm in}^i[\omega]+ \mathcal{G} \cdot \Xi_m[\omega]   \cdot \sqrt{\Gamma}  \cdot B_{\rm in}[\omega].
\end{multline}
Here we get an explicit expression for the coupling matrix as defined the main text:
\begin{equation}
T[\omega]\equiv - \mathcal{G} \cdot\Xi_m[\omega] \cdot  \mathcal{H}
\end{equation}
that contains coupling amplitudes for all phonon-mediated photon-photon couplings. Furthermore we  identify as in the main text the global photonic system susceptibility: \begin{equation}
\begin{array}{*3{>{\displaystyle}l}}
\chi[\omega] &\equiv& \Xi_c[\omega]^{-1} - \mathcal{G}  \cdot\Xi_m[\omega] \cdot \mathcal{H}\\
&=& \Xi_c[\omega]^{-1} + T[\omega]. 
\end{array}
\end{equation}
We also get the expression of the matrix $U[\omega]$ of the main text that characterizes cavity heating  due to mechanical thermal or quantum noise:
\begin{equation}
U[\omega]\equiv  \mathcal{G} \cdot\Xi_m[\omega] \cdot  \sqrt{\Gamma}
\end{equation}

\subsection{Phase-insensitive directional amplifier case}

In the present experimental case cavity 1 is only pumped on red sidebands so that $J_{1i}=0$ and cavity 2 only on blue sidebands so that $G_{2i}=0$, which yields the hollow  $T[\omega]$ coupling matrix reproduced in the main text, with coefficients:
\begin{equation}
\begin{array}{*3{>{\displaystyle}l}}
T_{11}[\omega] &=& |G_{11}|^2 \chi_{m,1}[\omega] + |G_{12}|^2 \chi_{m,2}[\omega]\\[10pt] 
T_{12}[\omega] &=& G_{11}J_{21}\; \chi_{m,1}[\omega] + G_{12}J_{22}\; \chi_{m,2}[\omega]\\[10pt]
T_{21}[\omega] &=& - \Big(G_{11}^*J_{21}^*\; \chi_{m,1}[\omega] + G_{12}^*J_{22}^*\; \chi_{m,2}[\omega]\Big)\\[10pt]
T_{22}[\omega] &=& -\Big(|J_{21}|^2 \chi_{m,1}[\omega] + |J_{22}|^2\chi_{m,2}[\omega]\Big)\\[10pt]
\end{array}
\end{equation}
which leads to the expressions given in the main text in terms of cooperativities, decay rates and pump relative phases.

\section{Amplifier parameters}

\subsection{$S_{\rm opt}$ elements}
$S_{\rm opt}$ has the same structure as the $T$ matrix, that is:
\begin{equation}
S_{\rm opt}[\omega]=\begin{pmatrix}
S_{11}[\omega] & 0 & 0 & S_{12}[\omega] \\
0& S_{11}[\omega]^*  S_{12}[\omega]^*& 0 & \\
0& S_{21}[\omega]^*  S_{22}[\omega]^*& 0 & \\
S_{21}[\omega] & 0 & 0 & S_{22}[\omega] 
\end{pmatrix}
\end{equation}
with the following coefficients:
\begin{equation}
\left\{\begin{array}{*4{>{\displaystyle}l}}
S_{11}[\omega]&=& 1-&\kappa_1^e\chi_{c,1}\;\frac{1+\chi_{c,2}T_{22}}{D}\\[10pt]
S_{12}[\omega]&=&&\sqrt{\kappa_1^e\kappa_2^e}\;\frac{\chi_{c,1}\chi_{c,2}T_{12}}{D}\\[10pt]
S_{21}[\omega]&=&&\sqrt{\kappa_1^e\kappa_2^e}\;\frac{\chi_{c,1}\chi_{c,2}T_{21}}{D}\\[10pt]
S_{22}[\omega]&=& 1-&\kappa_2^e\chi_{c,2}\;\frac{1+\chi_{c,1}T_{11}}{D}\\[10pt]
\end{array}\right.
\end{equation}
with the common denominator:
\begin{equation}
D=(1+\chi_{c,1}T_{11})(1+\chi_{c,2}T_{22})-\chi_{c,1}\chi_{c,2}T_{12}T_{21}.
\end{equation}

\subsection{Impedance matching}
The reflection coefficient is:
\begin{equation}
|S_{11}[0]|=1-\frac{2r_1}{\frac{2C_1}{(1+4\delta^2)}+1}
\end{equation} 
There is no reflection on the input cavity if this coefficient is 0, that is, for $r_1\neq \frac{1}{2}$ (input cavity not critically coupled, $\kappa_1^{e}\neq\kappa_1^{i}$):
\begin{equation}
\delta=\frac{1}{2}\sqrt{\frac{2C_1}{1-2r_1}-1}=\frac{\sqrt{2C_1\left(\frac{\kappa_1}{\kappa_1^{e}-\kappa_1^{i}}\right)-1}}{2}
\end{equation} 
For non-lossy cavities, one recovers the criterion from the proposal: $\delta = \frac{2C_1-1}{2}$ which is possible for $C_1>0.5$. However, this criterion can never be met if the input cavity is undercoupled $\kappa_1^{i}\leq\kappa_1^{e}$, and is only met for:
\begin{equation}
C_1>0.5\, \frac{\kappa_1^{e}-\kappa_1^{i}}{\kappa_1}
\end{equation}
in the general lossy case. This indicates that, unsurprisingly, as losses will require increased red sideband cooperativities, they will also require increased blue sideband cooperativities to observe some gain, so that gain is more difficult to obtain with lossy cavities if the impedance matching condition is to be met.

\subsection{Isolation}
The isolation condition $S_{12}[0]=0$ is achieved when $T_{12}[0]=0$, that is:
\begin{equation}
e^{i(\theta_{11}+\theta_{21}-\theta_{12}-\theta_{22})}\frac{\gamma_1}{\frac{\gamma_1}{2}-i\delta_1 }  =- \frac{\gamma_2}{\frac{\gamma_2}{2}-i\delta_2 } 
\end{equation}
Hence two conditions concerning the modulus and phase of the members of this equality. The condition on the modulus reduces to:
\begin{equation}
\frac{1}{4}+\frac{\delta_1^2}{\gamma_1^2}=\frac{1}{4}+\frac{\delta_2^2}{\gamma_2^2} \rightarrow \delta_1^2\gamma_2^2=\delta_2^2\gamma_1^2
\end{equation}
With the proposition \cite{Malz2018} $\delta_1=\delta \gamma_1$ and $\delta_2=-\delta \gamma_2$ and denoting $\Phi=\theta_{11}+\theta_{21}-\theta_{12}-\theta_{22}$, the previous phase equality is equivalent to:
\begin{equation}
\Phi  = {\rm Arg}\left( \frac{-1+2i\delta}{\;\,\;1+2i\delta } \right)
\end{equation}
Note that this condition can be regardless of cavities quality factors, and even that the isolation conditions on $\delta$ and $\Phi$ are independent on cavities qualities factor.



\subsection{Gain}

\begin{figure}[h]
\includegraphics[width=7cm]{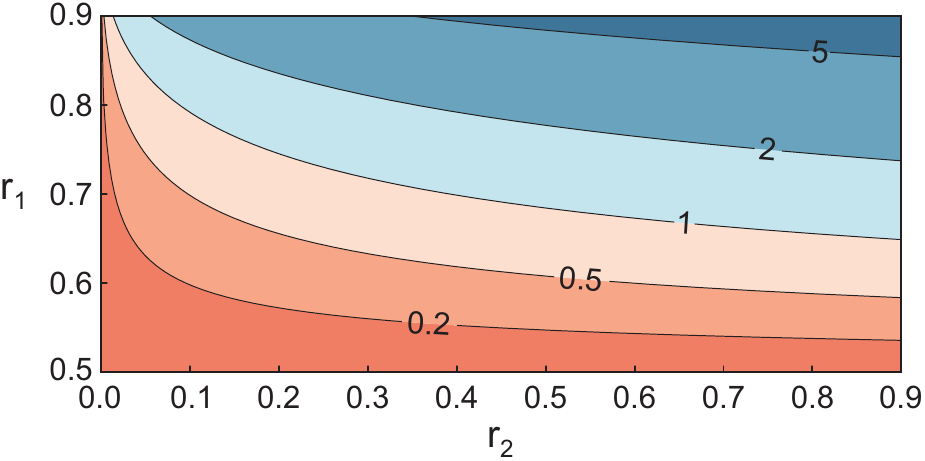}
\caption{Gain $|S_{21}[0]|$ in the limit of high cooperativities and in the impedance-matched case, as a function of $r_1=\frac{\kappa_1^e}{\kappa_1}$ and $r_2=\frac{\kappa_2^e}{\kappa_2}$.}
\label{fig:r1r2}
\end{figure}

The modulus of the gain of the amplifier is:
\begin{equation}\begin{array}{*3{>{\displaystyle}l}}
|S_{21}[0]|&=&\left|\frac{T_{21}\sqrt{\kappa_1^{e}\kappa_2^{e}}\chi_{c1}\chi_{c2} }{1+g_{11}\chi_{c1}+g_{22}\chi_{c2}+T_{11}T_{22}\chi_{c1}\chi_{c2} }\right|\\[10pt]
&=& \sqrt{r_1r_2}\frac{16 |\delta| \sqrt{C_1C_2(1+4\delta^2)}}{(2C_1+1+4\delta^2)|1+4\delta^2-2C_2|}
\end{array}
\end{equation} 
where we defined $r_j\equiv \frac{\kappa_j^{e}}{\kappa_j^{e}+\kappa_j^{i}}$.
The impedance-matching condition in the case of lossy input cavity is: $1+4\delta^2 = \frac{2C_1}{1-2r_1}$. The gain in the impedance-matched condition is then:
\begin{equation}
|S_{21}[0]|=\sqrt{r_1r_2}\left(\frac{2r_1-1}{r_1}\right)\frac{\sqrt{2C_2(2C_1+1-2r_1)}}{C_1+C_2(1-2r_1)}.
\end{equation}
In the limit $C_2=C_1+1$, $C_1\rightarrow\infty$, this gain is:
\begin{equation}
\lim_{C_1, C_2\rightarrow\infty}|S_{21}[0]|=\sqrt{\frac{r_2}{r_1}}\frac{1-2r_1}{r_1-1}.
\end{equation}
For $r_1>0.5$ (required by the impedance-matching condition) and $r_2>0$, this is a growing function of both $r_1$ and $r_2$, plotted on Figure \ref{fig:r1r2}. From this figure it becomes clear that the internal losses of the input cavity are much more deleterious to the amplifier gain these of the output cavity.

\section{Parasitic driving of mechanical oscillators}

In the case where the RSBD condition (Resolved SideBand Difference) is not completely achieved ($\Omega_2-\Omega_1\gtrsim \kappa_{1,2}$), some of the oscillating fields components present in equation \ref{eq:beforeRWA} that were neglected afterwards must be taken into account: those oscillating at $\pm (\Omega_2-\Omega_1)$. Then the following term to the expression of $b_1[\omega]$ and $b_2[\omega]$  from equation \ref{eq:aomega}:
\begin{equation}
\left\{\begin{array}{*3{>{\displaystyle}l}}
\Delta b_1[\omega] &=&  \chi_{m, 1}[\omega] \,i \sum_{i}\bigg((g_{0i1}\alpha_{i2-}^*)\,a_i[\omega-\Delta\Omega]+\\
&&\hfill(g_{0i1}\alpha_{i2+})\,a_i^\dagger[\omega-\Delta\Omega]\bigg)\\
\Delta b_2[\omega] &=&  \chi_{m, 2}[\omega] \,i \sum_{i}\bigg((g_{0i2}\alpha_{i1-}^*)\,a_i[\omega+\Delta\Omega]+\\
&&\hfill(g_{0i2}\alpha_{i1+})\,a_i^\dagger[\omega+\Delta\Omega]\bigg)\\[4pt]
\label{eq:bnoRWA}
\end{array}\right.
\end{equation}
where  $\Delta\Omega=(\Omega_1+\delta_1)-(\Omega_2+\delta_2)$, and the following term to the expression of $a_i[\omega]$ from equation \ref{eq:aomega}:
\begin{equation}
\begin{array}{*3{>{\displaystyle}l}}
\Delta a_i[\omega] =  \chi_{c, i}[\omega] \,i \bigg(  \!\!\!&&(g_{0i1}\alpha_{i2-})\,b_1[\omega+\Delta\Omega]\\
&+&(g_{0i1}\alpha_{i2+})\,b_1^\dagger[\omega-\Delta\Omega]\\[6pt]
&+&(g_{0i2}\alpha_{i1-})\,b_2[\omega-\Delta\Omega]\\
&+&(g_{0i2}\alpha_{i1+})\,b_2^\dagger[\omega+\Delta\Omega]\;\;\bigg)
\end{array}
\label{eq:anoRWA}
\end{equation}
Replacing the phononic operators in the photonic operators expressions yields expressions of $a_i[\omega]$ containing all $a_k[\omega]$,$a_k^\dagger[\omega]$, $a_k[\omega\pm\Delta\Omega]$,$a_k^\dagger[\omega\pm\Delta\Omega]$ and $a_k[\omega\pm2\Delta\Omega]$,$a_k^\dagger[\omega\pm2\Delta\Omega]$. We now determine what are the relevant generated terms if $\chi_{c,i}[\pm \Delta\Omega]$ is not neglected.

\subsection{First order perturbation from interaction with off-resonant cavity photons}
Let us temporarily use  the notation $A_n = \begin{pmatrix}a_1[\omega+n\Delta\Omega]&  a_1^\dagger[\omega+n\Delta\Omega]&a_2[\omega+n\Delta\Omega]&  a_2^\dagger[\omega+n\Delta\Omega]\end{pmatrix}^T$, for $n \in \mathbb{Z}$, and similar notations for all other frequency-dependent quantities. 
Under the RWAs, the expression of $A_n$ involves a coupling matrix $T_n$ itself involving the mechanical susceptibility $\Xi_{m,n}$. Now that one of the RWAs is dropped, it involves many more coupling matrices     containing mechanical susceptibilities $\Xi_{m,n}$ detuned by $n\Delta\Omega$, denoted $Q_n, R_n, S_n, V_n, W_n, X_n, Y_n, Z_n$ representing different coupling mechanisms between cavity fields \textit{via} mechanical oscillators, now that more of these mechanisms are driven. The expression of these matrices will be determined below. Momentarily omitting any input signal or noise, one can show the following expression:
\begin{equation}
\begin{array}{*3{>{\displaystyle}l}}
A_n = - \Xi_{c,n}\; & \Big(&Q_{n-1} \; A_{n-2}  \\
&+&( R_{n-1} + S_{n} )\;  A_{n-1}\\[5pt]
&+& (V_{n-1} + T_{n} + W_{n+1}) \; A_{n} \\[5pt]
&+& ( X_{n} + Y_{n+1} ) \;  A_{n+1}\\ [5pt]
&+& Z_{n+1}\;  A_{n+2} \Big).
\end{array}
\label{eq:recursionN}
\end{equation}
Since they all involve strongly filtering mechancial susceptibilities, all the coupling matrices can be neglected off-resonance: $Q_n...Z_n=0$ for $|n|>0$.  Writing equation \ref{eq:recursionN} for $n=0$, one then gets
\begin{equation}
A_0 = -\Xi_{c,0}  \bigg( T_{0} A_{0} +  S_{0}  A_{-1}  + X_{0} A_{1} \bigg).
\end{equation}
Injecting the expressions of $A_{\pm 1}$:
\begin{equation}
\begin{array}{*2{>{\displaystyle}l}}
A_0 = -\Xi_{c,0}  \bigg\{  &\Big(T_{0} + S_{0} \Xi_{c,-1} Y_{0} +  X_{0} \Xi_{c,1} R_{0} \Big)  A_{0}\\
&+  \Big( S_{0} \Xi_{c,-1} W_{0} + X_{0} \Xi_{c,1}  Q_{0} \Big) A_{-1}\\
&+ \Big(  S_{0} \Xi_{c,-1} Z_{0} +X_{0}\Xi_{c,1}V_{0}   \Big) A_{1} \bigg\}.
\end{array}
\end{equation}
All terms generated by the replacement of $A_{\pm 1}$ in this new expression will contain products of off-resonant cavity susceptibilities such as $\Xi_{c,\pm 1}\Xi_{c,\pm 1}$ which scales with $\kappa^2/\Delta\Omega^2$. They are therefore neglected as second order terms. The development is then truncated at:
\begin{equation}
A_0 \simeq -\Xi_{c,0}   \Big(T_{0} + S_{0}\Xi_{c,-1}Y_{0} +  X_{0}\Xi_{c,1}R_{0}\Big)   A_{0} .
\label{eq:devRWA}
\end{equation}
In first order in $\kappa/\Delta\Omega$, the coupling matrix around the resonance is modified by two terms representing interactions assisted by cavity photons from other manifolds. Note that these two coefficients count because e.g. fields $a_j[\omega]$ involving $b_j[\omega+\Delta\Omega]$ -- and therefore $\chi_{m,j}[\omega+\Delta\Omega]$ -- were invoked evaluated at the frequency $\omega-\Delta\Omega$ and finally contributed with $\chi_{m,j}[\omega]$, which is non-negligible at $\omega\simeq 0$. In other words, trips on other frequency manifolds $\omega\simeq\pm \Delta\Omega$ are allowed provided excitations come back from them to the $\omega\simeq0$ manifold.

\subsection{Explicit expressions of additional terms}
Equations \ref{eq:bnoRWA} and \ref{eq:anoRWA} lead to the matrix form:
\begin{equation}
\begin{array}{*3{>{\displaystyle}l}}
A_n&=& \Xi_{c,n} \Big(  \mathcal{G}   B_n \;+ \;\mathcal{G}^-   B_{n-1} \; + \;\mathcal{G}^+   B_{n+1}  \Big)\\
B_n &=&  \Xi_{m,n} \Big(  \mathcal{H}   A_n  + \mathcal{H}^-   A_{n-1}  + \mathcal{H}^+  A_{n+1}  \Big)
\end{array}
\end{equation}
with new coupling matrices $\mathcal{G}^\pm$ and  $\mathcal{H}^\pm$ that couple photons or phonons to the previous or next manifold of phonons or photons. We introduce $\tilde{G}_{ij}$ ($\tilde{J}_{ij}$) the multiphoton optomechanical coupling of MO $j$ to cavity $i$ enhanced by the red (blue) sideband intended for the other MO than $j$, e.g: $\tilde{G}_{11}=g_{011}\alpha_{12-}$. The expressions of the additional coupling matrices are:
\begin{equation}
\mathcal{G}^-\equiv -i \begin{pmatrix}
0 & \tilde{J}_{11} &  \tilde{G}_{12} &0\\
0 & -\tilde{G}_{11}^* &  -\tilde{J}_{12}^* &0\\
0 & \tilde{J}_{21} &  \tilde{G}_{22} &0\\
0 & -\tilde{G}_{21}^* &  -\tilde{J}_{22}^* &0\\
\end{pmatrix}
\end{equation}
\begin{equation}
\mathcal{G}^+\equiv -i \begin{pmatrix}
\tilde{G}_{11} & 0 & 0 & \tilde{J}_{12} \\
-\tilde{J}_{11}^* & 0 &0 & -\tilde{G}_{12}^* \\
\tilde{G}_{21} & 0 & 0 & \tilde{J}_{22} \\
-\tilde{J}_{21}^* & 0 &0 & -\tilde{G}_{22}^* \\
\end{pmatrix}
\end{equation}
\begin{equation}
\mathcal{H}^-\equiv -i \begin{pmatrix}
\tilde{G}_{11}^* & \tilde{J}_{11} & \tilde{G}_{21}^* & \tilde{J}_{21} \\
0 & 0 & 0 & 0\\
0 & 0 & 0 & 0\\
-\tilde{J}_{12}^* & -\tilde{G}_{12} & -\tilde{J}_{22}^* & -\tilde{G}_{22} \\
\end{pmatrix}
\end{equation}
\begin{equation}
\mathcal{H}^+\equiv  -i \begin{pmatrix}
0 & 0 & 0 & 0\\
-\tilde{J}_{11}^* & -\tilde{G}_{11} & -\tilde{J}_{21}^* & -\tilde{G}_{21} \\
\tilde{G}_{12}^* & \tilde{J}_{12} & \tilde{G}_{22}^* & \tilde{J}_{22} \\
0 & 0 & 0 & 0\\
\end{pmatrix}
\end{equation}
Replacing the phononic operators by their expression in the photonic operators expression, one identifies the expressions of coefficients $Q_0...Z_0$. Among these, the interesting ones according to equation \ref{eq:devRWA} are:
\begin{equation}
\begin{array}{*7{>{\displaystyle}l}}
S_0 &= &  -\mathcal{G} \cdot \Xi_m[\omega] \cdot  \mathcal{H}^-, & \qquad &Y_0&=&  -\mathcal{G}^+ \cdot \Xi_m[\omega] \cdot  \mathcal{H}  \\[4pt]
X_0 &= &  -\mathcal{G} \cdot \Xi_m[\omega] \cdot  \mathcal{H}^+, & \qquad &R_0&=&  -\mathcal{G}^- \cdot \Xi_m[\omega] \cdot  \mathcal{H}  
\end{array}
\end{equation}
Note that both additional terms: $S_{0}\Xi_{c,-1}Y_{0} $ and $ X_{0}\Xi_{c,1}R_{0}$ to the coupling matrix $T$ have the same zero-elements as $T$: therefore the phase-conjugating nature of the frequency converting is not modified when taking this first-order perturbation into account.

\subsection{Application to the case of the directional amplifier}

\begin{figure}[h]
\includegraphics[width=\columnwidth]{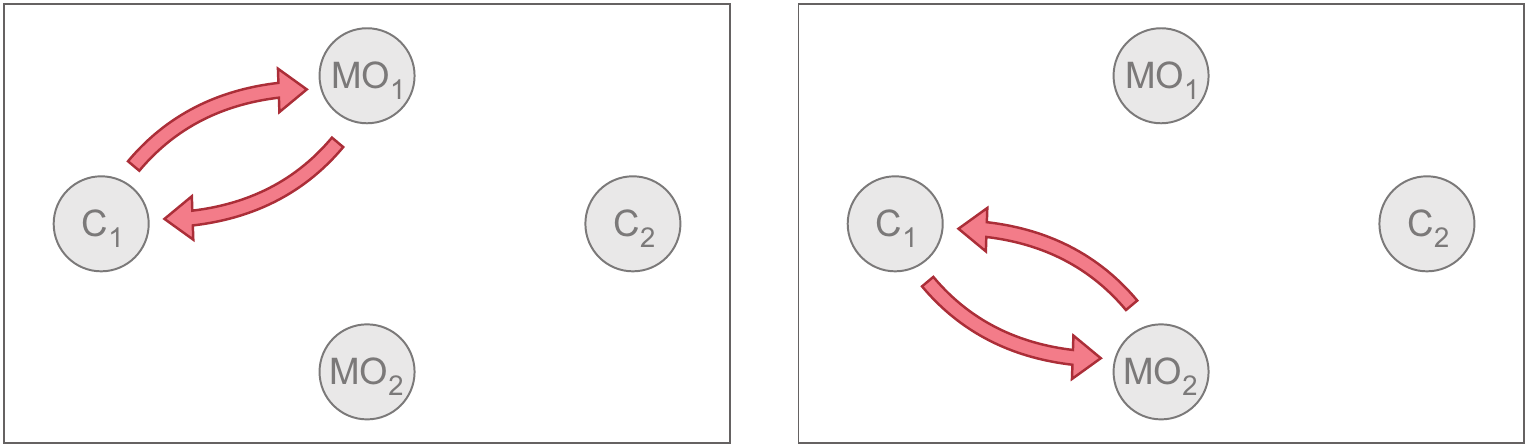}
\caption{RWA-contributions to $T_{11}$}
\label{fig:T11RWA}
\end{figure}

In the case of the directional amplifier studied here, the self-coupling coefficient $T_{11}[\omega]$ had two terms in the RWA framework:
\begin{equation}
T_{11}[\omega]= \frac{C_1\kappa_1}{4}\Big(\gamma_1 \chi_{m,1}[\omega] + \gamma_2 \chi_{m,2}[\omega]\Big)
\end{equation}
which can be represented by the graphs of \fref{fig:T11RWA}.

\begin{figure}[h]
\includegraphics[width=\columnwidth]{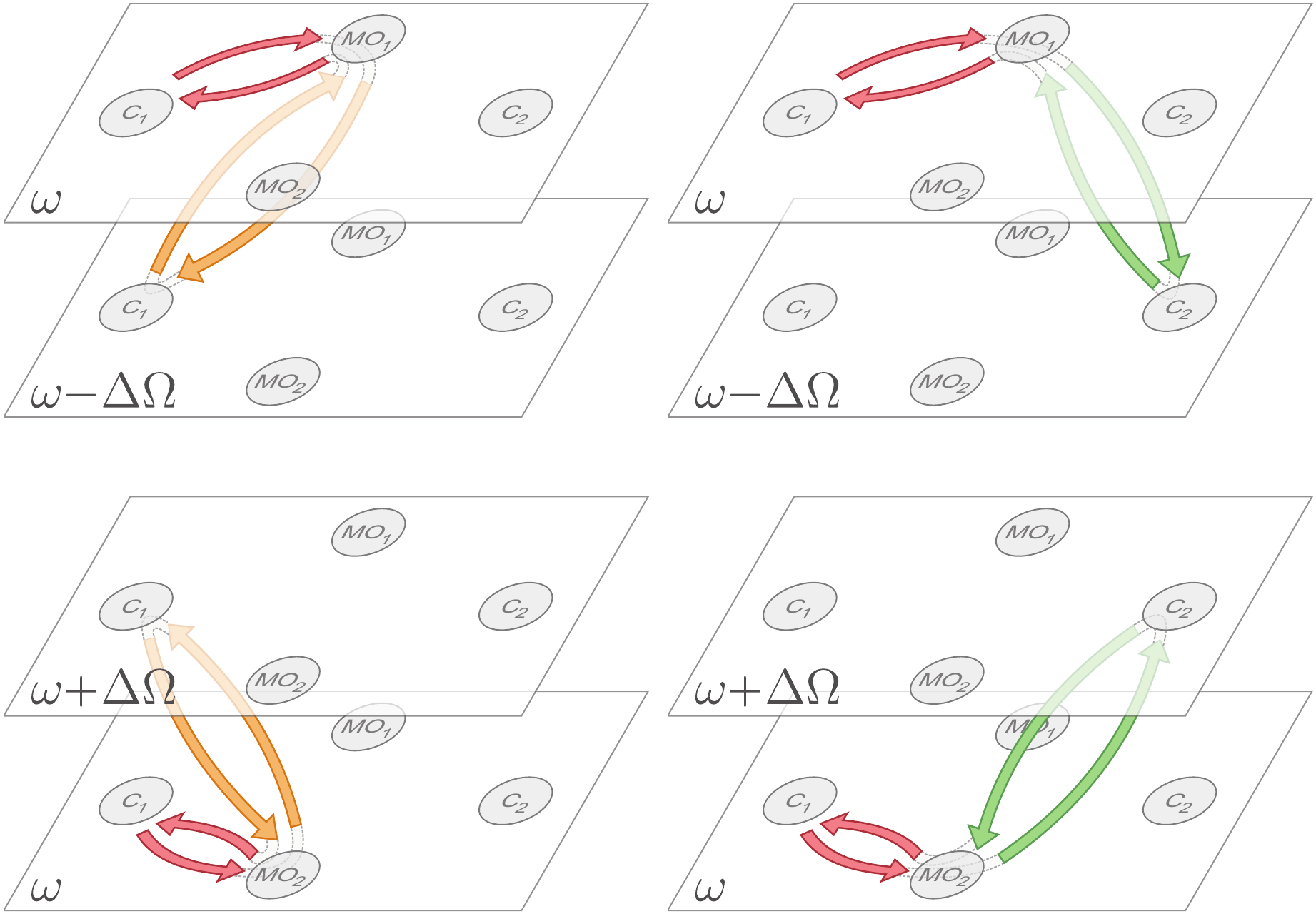}
\caption{Perturbations to $T_{11}$ to first order in $\kappa/\Delta\Omega$}
\label{fig:T11noRWA}
\end{figure}

The first order perturbation adds the following terms to $T_{11}[\omega]$:
\begin{equation}
\begin{array}{*4{>{\displaystyle}l}}
T_{11}[\omega]\rightarrow T_{11}[\omega]&+&\chi_{m,1}^2[\omega] \Big(&|G_{11}\tilde{G}_{11}|^2\chi_{c,1}[\omega-\Delta\Omega]\\
&&&\!\!\!\!-  |G_{11}\tilde{J}_{21}|^2\chi_{c,2}^*[\omega-\Delta\Omega] \,\Big)\\[8pt]
&+&\chi_{m,2}^2[\omega] \Big(&|G_{12}\tilde{G}_{12}|^2\chi_{c,1}[\omega+\Delta\Omega] \\
&&&\!\!\!\!-  |G_{12}\tilde{J}_{22}|^2\chi_{c,2}^*[\omega+\Delta\Omega] \,\Big)\\[5pt]
\end{array}
\end{equation}
which are represented on several-manifolds-graphs as shown on \fref{fig:T11noRWA}.

These additional contributions are in fact the result of the dynamical backaction due to the off-resonant terms from each cavities on each MOs:
\begin{equation}
\begin{array}{*5{>{\displaystyle}l}}
\chi_{m,1}[\omega]&\rightarrow& \chi_{m,1}[\omega] \Big(1&+&\chi_{m,1}[\omega]|\tilde{G}_{11}|^2\chi_{c,1}[\omega-\Delta\Omega]\\
&&&-&\chi_{m,1}[\omega]|\tilde{J}_{21}|^2\chi_{c,2}^*[\omega-\Delta\Omega]\Big)\\[5pt]
\chi_{m,2}[\omega]&\rightarrow& \chi_{m,1}[\omega] \Big(1&+&\chi_{m,2}[\omega]|\tilde{G}_{12}|^2\chi_{c,1}[\omega+\Delta\Omega]\\
&&&-&\chi_{m,2}[\omega]|\tilde{J}_{22}|^2\chi_{c,2}^*[\omega+\Delta\Omega]\Big)
\end{array}
\end{equation}
Interestingly, the device's behavior is, in the RWA framework, governed by the \textit{bare} mechanical susceptibilities, that is, it is insensitive to backaction from the pump tones that drive each MO. Only the backaction due to tones not intended to drive each mode matters to the amplifier's quality (to first order in $\kappa/\Delta\Omega$.) In terms of experimental parameters, the additional terms to $T_{11}$ are:
\begin{equation}
\begin{array}{*3{>{\displaystyle}l}}
\;\;\;\;\gamma_1\gamma_2\chi_{m,1}^2[\omega] \Big\{&&\left(\frac{g_{011}}{g_{012}}\right)^2 \frac{C_1^2\kappa_1^2}{16} \chi_{c,1}[\omega-\Delta\Omega]\\
&-& \left(\frac{g_{021}}{g_{022}}\right)^2 \frac{C_2^2\kappa_2^2}{16}  \chi_{c,2}^*[\omega-\Delta\Omega] \,\Big\}\\[10pt]
+ \;\gamma_1\gamma_2\chi_{m,2}^2[\omega] \Big\{&&\left(\frac{g_{012}}{g_{011}}\right)^2 \frac{C_1^2\kappa_1^2}{16} \chi_{c,1}[\omega+\Delta\Omega] \\
&-&  \left(\frac{g_{022}}{g_{021}}\right)^2 \frac{C_2^2\kappa_2^2}{16}\chi_{c,2}^*[\omega+\Delta\Omega] \,\Big\}\end{array}
\end{equation}

\begin{figure}[h]
\includegraphics[width=\columnwidth]{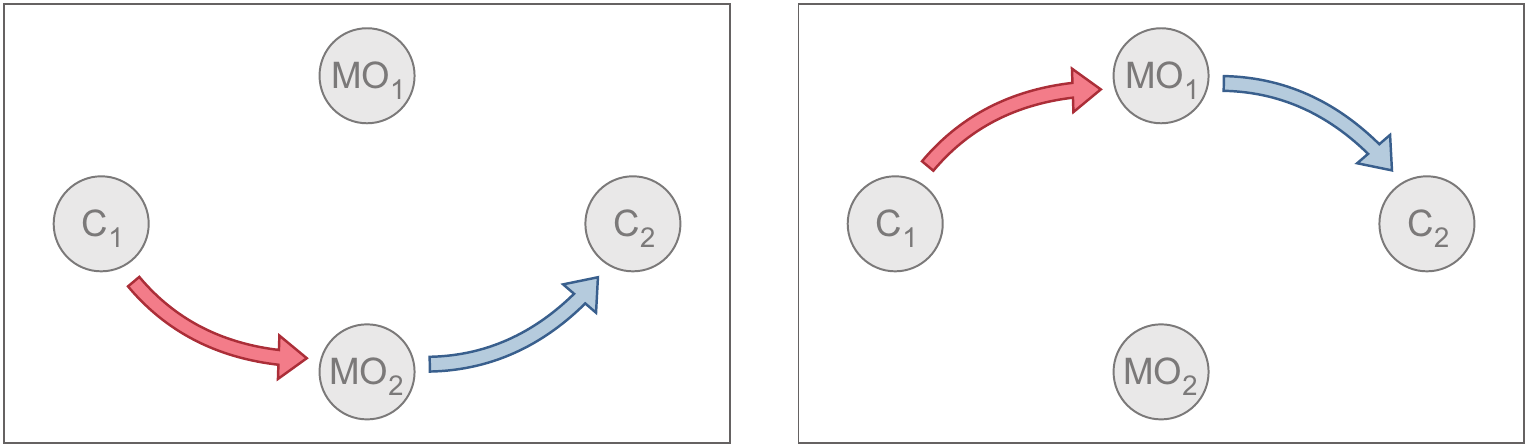}
\caption{RWA-contributions to $T_{12}$}
\label{fig:T12RWA}
\end{figure}

In the same idea, the zero-order expression of $T_{12}[\omega]$ involves two terms (made to interfere destructively):
\begin{equation}
T_{12}[\omega] = \frac{\sqrt{C_1C_2\kappa_1\kappa_2}}{4} \Big(e^{i\Phi/2}\gamma_1 \chi_{m,1}[\omega] + e^{-i\Phi/2}\gamma_2 \chi_{m,2}[\omega]\Big)
\end{equation}  
that can be represented as on \fref{fig:T12RWA}.

\begin{figure}[h]
\includegraphics[width=\columnwidth]{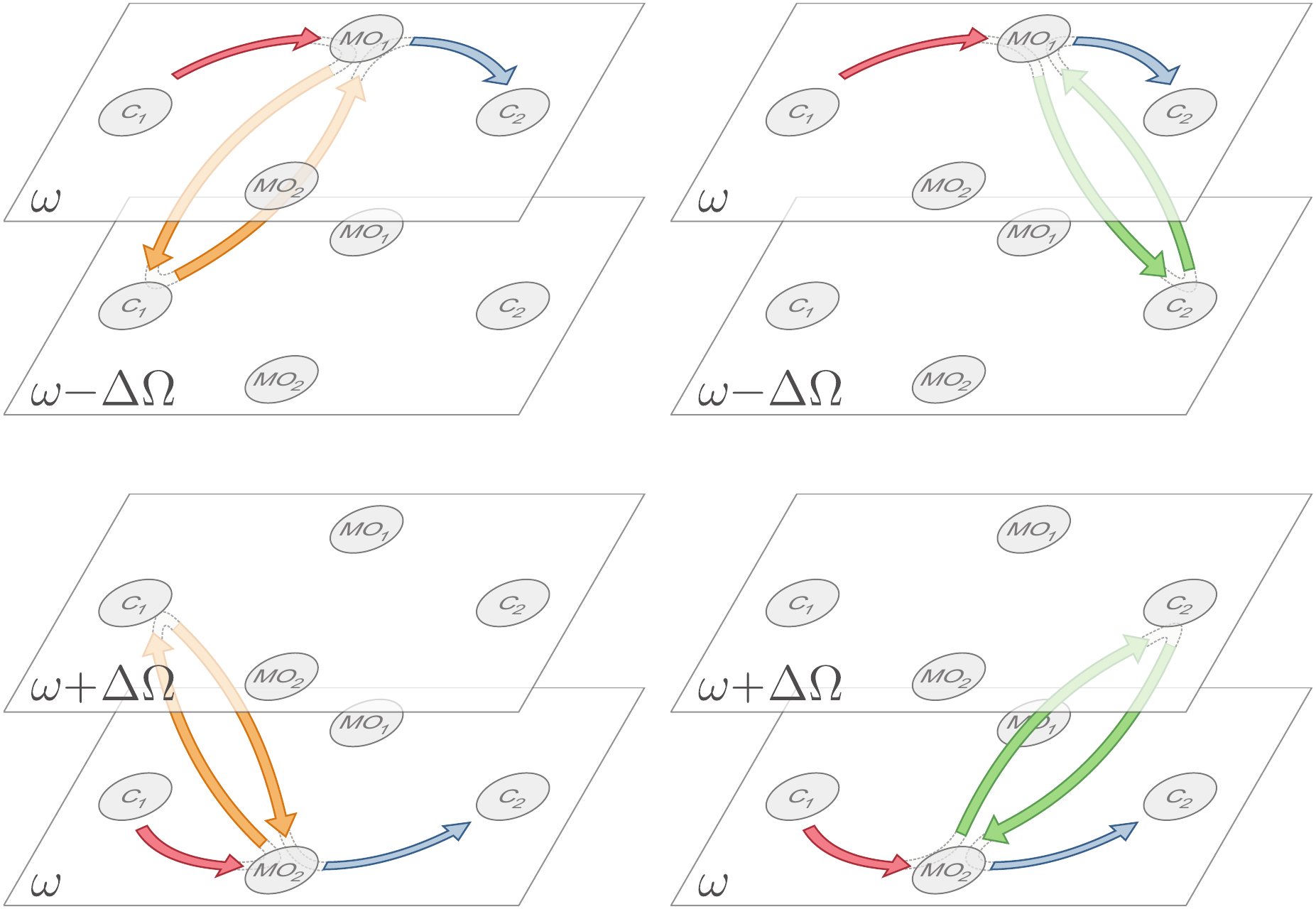}
\caption{Perturbations to $T_{12}$ to first order in $\kappa/\Delta\Omega$}
\label{fig:T12noRWA}
\end{figure}

The first order perturbation adds the following terms:
\begin{equation}
\begin{array}{*4{>{\displaystyle}l}}
T_{12}[\omega]\rightarrow T_{12}[\omega]\!\!&+\!\!&\chi_{m,1}^2[\omega] \Big(&\,|\tilde{G}_{11}|^2G_{11}J_{21}\chi_{c,1}[\omega-\Delta\Omega]\\
&&&\!\!\!\!- |\tilde{J}_{21}|^2G_{11}J_{21}\chi_{c,2}^*[\omega-\Delta\Omega] \,\Big)\\[8pt]
&+\!\!&\chi_{m,2}^2[\omega] \Big(&\,|\tilde{G}_{12}|^2G_{12}J_{22}\chi_{c,1}[\omega+\Delta\Omega]\\
&&&\!\!\!\!-  |\tilde{J}_{22}|^2G_{12}J_{22}\chi_{c,2}^*[\omega+\Delta\Omega] \,\Big)
\end{array}
\end{equation}
that are shown in graph representation on \fref{fig:T12noRWA}.

In terms of experimental parameters, the additional terms are:
\begin{multline}
\frac{\gamma_1\gamma_2 \sqrt{C_1C_2\kappa_2\kappa_1} }{16} \times\\
\Bigg\{\chi_{m,1}^2[\omega] e^{i\Phi/2} \Bigg[\,\left(\frac{g_{011}}{g_{012}}\right)^2 \!\!\sqrt{C_1\kappa_1} \chi_{c,1}[\omega-\Delta\Omega] \\
\hfill - \left(\frac{g_{021}}{g_{022}}\right)^2 \sqrt{C_2\kappa_2}  \chi_{c,2}^*[\omega-\Delta\Omega] \,\Bigg] \\
+ \chi_{m,2}^2[\omega]e^{-i\Phi/2}  \Bigg[\,\left(\frac{g_{012}}{g_{011}}\right)^2 \sqrt{C_1\kappa_1} \chi_{c,1}[\omega+\Delta\Omega] \\
\hfill -   \left(\frac{g_{022}}{g_{021}}\right)^2 \sqrt{C_2\kappa_2}\chi_{c,2}^*[\omega+\Delta\Omega] \,\Bigg]\Bigg\} .
\end{multline}
Notably, only the phase $\Phi$ appears in these terms. Furthermore, the off-resonant cavity susceptibilities have real parts of different signs for backaction on MO1 and MO2: pump conditions calibrated for the RWA-framework terms to interfere destructively will not result in a destructive interference of the terms developed in this paragraph.

To summarize, we showed that, at first order of perturbation in $\kappa/\Delta\Omega$:
\begin{itemize} \item{each pump tone does not only address one mechanical mode only but induces a dynamical backaction on the second which has consequences on the device (unlike the backaction of tones present within the RWA picture which have no effect on the amplifier). The exact form of the effect is strongly dependent on the ratio of single-photon couplings, and generally influences the working point and bandwidth,}
\item{the phase-insensitive (phase conjugating) nature of the device is conserved,}
\item{only one pump phase appears as in the RWA picture, so that, for example, there is no way to cancel out parasitic effects by tuning additional phase degrees of freedom,}
\item{the ideal working point of the device will be modified and its quality may be altered since first-order terms do not interfere in the same way as zero-order terms. However, the additional dynamical backaction may happen to increase the device bandwidth if the dressed mechanical resonances are broadened. This has been observed in Ref.~\cite{Bernier2017} where only red tones were used, but is not necessarily the case here since the modes are subject to backaction of different signs from the red and the blue sidebands.}
\end{itemize}

\section{Noise subtraction}

The network analyzer (NA) measures a ratio of input to output powers $S_{ij,\rm meas}[\omega]=P_{{\rm in},i}[\omega]/P_{{\rm out},j}$ where $i,j$ denote the frequency ranges around cavities $i,j$ (note that  here input and output denominations were chosen with respect to the measurement instrument and not the device). Along with the power yielded from the amplifier excitation, the power corresponding to noise integration on the NA bandwidth $BW$ ($BW=20\,\rm Hz$) also contributes to $P_{{\rm in},i}$. The noise is therefore independently measured on a spectrum analyzer (SA) that gives the power $P_{\rm noise}$ integrated on the SA resolution bandwidth $RBW$ ($RBW=30\,\rm Hz$). Subsequently this noise is removed from the measured S-parameters, taking into account the different integration bandwidths, to yield the pure transmission coefficients:
\begin{equation}
S_{ij}[\omega]=S_{ij,\rm meas}[\omega]-\frac{BW}{RBW}\frac{P_{{\rm noise},i}[\omega]}{P_{{\rm out},j}}.
\end{equation}

\section{System gain calibration}
After noise subtraction, the transfer parameters of the amplifier are extracted from the measured signals, that is, these signals are corrected by a factor taking into account to the total attenuation and amplification from cables, amplifiers and other  elements on the input and output lines of the cryostat.\\

The input and output sides of these measurement lines show a frequency dependence. The four transfer parameters are measured by pumping and probing very different frequency ranges around the two resonances of the superconducting circuit ($5.6\,\rm GHz$ and $3.9\,\rm GHz$), therefore they are affected by different gains. More precisely, the raw measured transfer coefficients are related to the device transfer coefficients by:
\begin{equation}
\begin{array}{*3{>{\displaystyle}l}}
S_{11}^{\rm raw}&=& \eta_{1}^{\rm in}\;\eta_{1}^{\rm out}\; S_{11}\\[6pt]
S_{12}^{\rm raw}&=& \eta_{2}^{\rm in}\;\eta_{1}^{\rm out}\; S_{12}\\[6pt]
S_{21}^{\rm raw}&=& \eta_{1}^{\rm in}\;\eta_{2}^{\rm out}\; S_{21}\\[6pt]
S_{22}^{\rm raw}&=& \eta_{2}^{\rm in}\;\eta_{2}^{\rm out}\; S_{22}
\end{array}
\end{equation}
where $\eta_{\rm in}$ are pump efficiencies characterizing the total attenuation on the input side, $\eta_{\rm out}$ are measurement efficiencies denoting the total amplification on the output side and the indices 1,2 represent the two ranges of frequencies (around cavity 1 and 2). These gains are assumed not to vary significantly on each of these frequency ranges, which seems a very reasonable assumption from their typical frequency dependence out of cavity resonance.\\

\begin{figure}
\includegraphics[width=0.9\columnwidth]{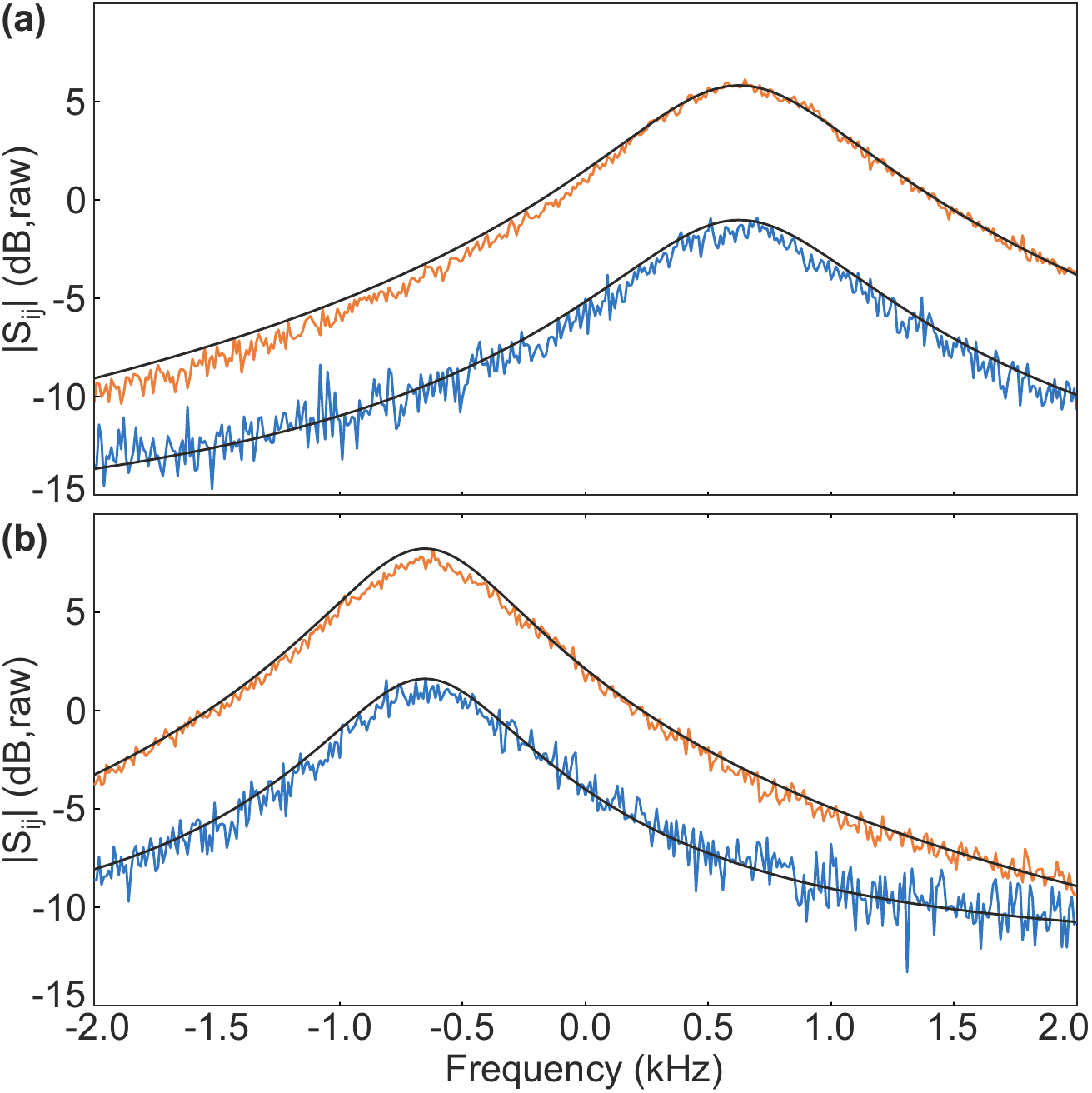}
\caption{\emph{Single-MO amplifiers used for cross-gain calibration.}  (a) Single-MO amplifier built with MO1 $|S_{12}|$ (orange) and  $|S_{21}|$ (blue) parameters corrected by the average gain $\sqrt{\eta_{2}^{\rm in}\;\eta_{1}^{\rm out} \eta_{1}^{\rm in}\;\eta_{2}^{\rm out}}$, still showing a $6.8\,\rm dB$ imbalance. This value was calibrated from the fit in black lines. A constant background was added in the fit expression to reproduce the noise floor. (b) Same data taken for the other single-MO amplifier built with MO2, showing the same imbalance of $6.8\,\m{dB}$, obtained from a separate fit.}
\label{calib}
\end{figure}

The two measurement system gains affecting $S_{11}$ and $S_{22}$ are easily measured from the response of the device out of, but close to, the cavity resonances, as explained in the article: $20\,{\rm log}_{10} ( \eta_{1}^{\rm in}\,\eta_{1}^{\rm out})=-61.2 \pm 0.1\,\rm dB$ and $20\,{\rm log}_{10} (\eta_{2}^{\rm in}\,\eta_{2}^{\rm out})=-58.8 \pm 0.2\,\rm dB$. The uncertainties represent the typical standard deviations of $S_{11}$ and $S_{22}$  out of resonance.

However, the two other gains, that affect $S_{12}$ and $S_{21}$, cannot be estimated in the same way since there is no frequency-converting transfer out of  cavity resonances. The product of these transmission gains is however known (to be equal to the product of the two reflection gains):
\begin{equation}
20\,{\rm log}_{10} \left(  \eta_{2}^{\rm in}\;\eta_{1}^{\rm out} \times \eta_{1}^{\rm in}\;\eta_{2}^{\rm out} \right) = -120.0\pm 0.3\,\rm dB.
\end{equation}\\
Single-mechanical-oscillators-amplifiers are reciprocal \cite{CasparAmp} $S_{12}=S_{21}$, but we measure from both amplifiers built with each of the mechanical oscillators:
\begin{equation}
\begin{array}{*3{>{\displaystyle}l}}
20\,{\rm log}_{10} \left(|S_{12}^{\rm raw}/S_{21}^{\rm raw}| \right)&=&20\,{\rm log}_{10} \left( \eta_{2}^{\rm in}\;\eta_{1}^{\rm out} /\eta_{1}^{\rm in}\;\eta_{2}^{\rm out}  \right)\\[4pt]
&=& 6.8 \pm 0.8\,\rm dB.
\end{array}
\end{equation}
The uncertainty corresponds to the sum of the standard deviations of $S_{12}$ and $S_{21}$ data for 1-MO amplifiers from which the fit value was removed.  The remaining gains can be computed from this imbalance:
\begin{equation}
\left\{\begin{array}{*3{>{\displaystyle}l}}
20\,{\rm log}_{10} \left(\eta_{2}^{\rm in}\;\eta_{1}^{\rm out}\right)&=& -56.6\pm 1.1\,\rm dB\\[4pt]
20\,{\rm log}_{10} \left(\eta_{1}^{\rm in}\;\eta_{2}^{\rm out}\right)&=& -63.4\pm 1.1\,\rm dB.
\end{array}\right.
\end{equation}

This calibration 
entails an input efficiency difference of $20\,{\rm log}_{10} \left(\eta_{1}^{\rm in}/\eta_{2}^{\rm in}\right)=-4.6\,\rm dB$ between cavity 1 ($5.6\,\rm GHz$) and cavity 2 ($3.9\,\rm GHz$) frequency ranges. This is probably due to the use of resistive lines on this side of the measurement system, which display higher losses at larger frequencies.
On the other hand, the gain discrepancy has the other sign on the output side $20\,{\rm log}_{10} \left(\eta_{1}^{\rm out}/\eta_{2}^{\rm out}\right)=2.2\,\rm dB$. An independent calibration of the output lines confirmed the $2.2\pm 0.3\,\rm dB$ lower gain at the lower frequencies of cavity 2 ($3.9\,\rm GHz$) compared to cavity 1 frequencies  ($5.6\,\rm GHz$), mainly due to the proximity to the cut frequency of a high-pass filter ($4\,\rm GHz$).


\begin{thebibliography}{42}%
\makeatletter
\providecommand \@ifxundefined [1]{%
 \@ifx{#1\undefined}
}%
\providecommand \@ifnum [1]{%
 \ifnum #1\expandafter \@firstoftwo
 \else \expandafter \@secondoftwo
 \fi
}%
\providecommand \@ifx [1]{%
 \ifx #1\expandafter \@firstoftwo
 \else \expandafter \@secondoftwo
 \fi
}%
\providecommand \natexlab [1]{#1}%
\providecommand \enquote  [1]{``#1''}%
\providecommand \bibnamefont  [1]{#1}%
\providecommand \bibfnamefont [1]{#1}%
\providecommand \citenamefont [1]{#1}%
\providecommand \href@noop [0]{\@secondoftwo}%
\providecommand \href [0]{\begingroup \@sanitize@url \@href}%
\providecommand \@href[1]{\@@startlink{#1}\@@href}%
\providecommand \@@href[1]{\endgroup#1\@@endlink}%
\providecommand \@sanitize@url [0]{\catcode `\\12\catcode `\$12\catcode
  `\&12\catcode `\#12\catcode `\^12\catcode `\_12\catcode `\%12\relax}%
\providecommand \@@startlink[1]{}%
\providecommand \@@endlink[0]{}%
\providecommand \url  [0]{\begingroup\@sanitize@url \@url }%
\providecommand \@url [1]{\endgroup\@href {#1}{\urlprefix }}%
\providecommand \urlprefix  [0]{URL }%
\providecommand \Eprint [0]{\href }%
\providecommand \doibase [0]{http://dx.doi.org/}%
\providecommand \selectlanguage [0]{\@gobble}%
\providecommand \bibinfo  [0]{\@secondoftwo}%
\providecommand \bibfield  [0]{\@secondoftwo}%
\providecommand \translation [1]{[#1]}%
\providecommand \BibitemOpen [0]{}%
\providecommand \bibitemStop [0]{}%
\providecommand \bibitemNoStop [0]{.\EOS\space}%
\providecommand \EOS [0]{\spacefactor3000\relax}%
\providecommand \BibitemShut  [1]{\csname bibitem#1\endcsname}%
\let\auto@bib@innerbib\@empty
\bibitem [{\citenamefont {Abdo}\ \emph {et~al.}(2013)\citenamefont {Abdo},
  \citenamefont {Sliwa}, \citenamefont {Frunzio},\ and\ \citenamefont
  {Devoret}}]{Abdo2013}%
  \BibitemOpen
  \bibfield  {author} {\bibinfo {author} {\bibfnamefont {B.}~\bibnamefont
  {Abdo}}, \bibinfo {author} {\bibfnamefont {K.}~\bibnamefont {Sliwa}},
  \bibinfo {author} {\bibfnamefont {L.}~\bibnamefont {Frunzio}}, \ and\
  \bibinfo {author} {\bibfnamefont {M.}~\bibnamefont {Devoret}},\ }\bibfield
  {title} {\enquote {\bibinfo {title} {{Directional Amplification with a
  Josephson Circuit}},}\ }\href@noop {} {\bibfield  {journal} {\bibinfo
  {journal} {Phys. Rev. X}\ }\textbf {\bibinfo {volume} {3}},\ \bibinfo {pages}
  {031001} (\bibinfo {year} {2013})}\BibitemShut {NoStop}%
\bibitem [{\citenamefont {Abdo}\ \emph {et~al.}(2014)\citenamefont {Abdo},
  \citenamefont {Sliwa}, \citenamefont {Shankar}, \citenamefont {Hatridge},
  \citenamefont {Frunzio}, \citenamefont {Schoelkopf},\ and\ \citenamefont
  {Devoret}}]{Abdo2014}%
  \BibitemOpen
  \bibfield  {author} {\bibinfo {author} {\bibfnamefont {B.}~\bibnamefont
  {Abdo}}, \bibinfo {author} {\bibfnamefont {K.}~\bibnamefont {Sliwa}},
  \bibinfo {author} {\bibfnamefont {S.}~\bibnamefont {Shankar}}, \bibinfo
  {author} {\bibfnamefont {M.}~\bibnamefont {Hatridge}}, \bibinfo {author}
  {\bibfnamefont {L.}~\bibnamefont {Frunzio}}, \bibinfo {author} {\bibfnamefont
  {R.}~\bibnamefont {Schoelkopf}}, \ and\ \bibinfo {author} {\bibfnamefont
  {M.}~\bibnamefont {Devoret}},\ }\bibfield  {title} {\enquote {\bibinfo
  {title} {{Josephson Directional Amplifier for Quantum Measurement of
  Superconducting Circuits}},}\ }\href {\doibase
  10.1103/PhysRevLett.112.167701} {\bibfield  {journal} {\bibinfo  {journal}
  {Phys. Rev. Lett.}\ }\textbf {\bibinfo {volume} {112}},\ \bibinfo {pages}
  {167701} (\bibinfo {year} {2014})}\BibitemShut {NoStop}%
\bibitem [{\citenamefont {Kamal}\ \emph {et~al.}(2014)\citenamefont {Kamal},
  \citenamefont {Roy}, \citenamefont {Clarke},\ and\ \citenamefont
  {Devoret}}]{Kamal2014}%
  \BibitemOpen
  \bibfield  {author} {\bibinfo {author} {\bibfnamefont {A.}~\bibnamefont
  {Kamal}}, \bibinfo {author} {\bibfnamefont {A.}~\bibnamefont {Roy}}, \bibinfo
  {author} {\bibfnamefont {J.}~\bibnamefont {Clarke}}, \ and\ \bibinfo {author}
  {\bibfnamefont {M.~H.}\ \bibnamefont {Devoret}},\ }\bibfield  {title}
  {\enquote {\bibinfo {title} {{Asymmetric Frequency Conversion in Nonlinear
  Systems Driven by a Biharmonic Pump}},}\ }\href@noop {} {\bibfield  {journal}
  {\bibinfo  {journal} {Phys. Rev. Lett.}\ }\textbf {\bibinfo {volume} {113}},\
  \bibinfo {pages} {247003} (\bibinfo {year} {2014})}\BibitemShut {NoStop}%
\bibitem [{\citenamefont {Sliwa}\ \emph {et~al.}(2015)\citenamefont {Sliwa},
  \citenamefont {Hatridge}, \citenamefont {Narla}, \citenamefont {Shankar},
  \citenamefont {Frunzio}, \citenamefont {Schoelkopf},\ and\ \citenamefont
  {Devoret}}]{Sliwa2015}%
  \BibitemOpen
  \bibfield  {author} {\bibinfo {author} {\bibfnamefont {K.~M.}\ \bibnamefont
  {Sliwa}}, \bibinfo {author} {\bibfnamefont {M.}~\bibnamefont {Hatridge}},
  \bibinfo {author} {\bibfnamefont {A.}~\bibnamefont {Narla}}, \bibinfo
  {author} {\bibfnamefont {S.}~\bibnamefont {Shankar}}, \bibinfo {author}
  {\bibfnamefont {L.}~\bibnamefont {Frunzio}}, \bibinfo {author} {\bibfnamefont
  {R.~J.}\ \bibnamefont {Schoelkopf}}, \ and\ \bibinfo {author} {\bibfnamefont
  {M.~H.}\ \bibnamefont {Devoret}},\ }\bibfield  {title} {\enquote {\bibinfo
  {title} {{Reconfigurable Josephson Circulator/Directional Amplifier}},}\
  }\href@noop {} {\bibfield  {journal} {\bibinfo  {journal} {Phys. Rev. X}\
  }\textbf {\bibinfo {volume} {5}},\ \bibinfo {pages} {041020} (\bibinfo {year}
  {2015})}\BibitemShut {NoStop}%
\bibitem [{\citenamefont {Lecocq}\ \emph {et~al.}(2017)\citenamefont {Lecocq},
  \citenamefont {Ranzani}, \citenamefont {Peterson}, \citenamefont {Cicak},
  \citenamefont {Simmonds}, \citenamefont {Teufel},\ and\ \citenamefont
  {Aumentado}}]{Lecocq2017}%
  \BibitemOpen
  \bibfield  {author} {\bibinfo {author} {\bibfnamefont {F.}~\bibnamefont
  {Lecocq}}, \bibinfo {author} {\bibfnamefont {L.}~\bibnamefont {Ranzani}},
  \bibinfo {author} {\bibfnamefont {G.~A.}\ \bibnamefont {Peterson}}, \bibinfo
  {author} {\bibfnamefont {K.}~\bibnamefont {Cicak}}, \bibinfo {author}
  {\bibfnamefont {R.~W.}\ \bibnamefont {Simmonds}}, \bibinfo {author}
  {\bibfnamefont {J.~D.}\ \bibnamefont {Teufel}}, \ and\ \bibinfo {author}
  {\bibfnamefont {J.}~\bibnamefont {Aumentado}},\ }\bibfield  {title} {\enquote
  {\bibinfo {title} {{Nonreciprocal Microwave Signal Processing with a
  Field-Programmable Josephson Amplifier}},}\ }\href@noop {} {\bibfield
  {journal} {\bibinfo  {journal} {Phys. Rev. Applied}\ }\textbf {\bibinfo
  {volume} {7}},\ \bibinfo {pages} {024028} (\bibinfo {year}
  {2017})}\BibitemShut {NoStop}%
\bibitem [{\citenamefont {Westig}\ and\ \citenamefont
  {Klapwijk}(2018)}]{Westig2018}%
  \BibitemOpen
  \bibfield  {author} {\bibinfo {author} {\bibfnamefont {M.~P.}\ \bibnamefont
  {Westig}}\ and\ \bibinfo {author} {\bibfnamefont {T.~M.}\ \bibnamefont
  {Klapwijk}},\ }\bibfield  {title} {\enquote {\bibinfo {title} {{Josephson
  Parametric Reflection Amplifier with Integrated Directionality}},}\
  }\href@noop {} {\bibfield  {journal} {\bibinfo  {journal} {Phys. Rev.
  Applied}\ }\textbf {\bibinfo {volume} {9}},\ \bibinfo {pages} {064010}
  (\bibinfo {year} {2018})}\BibitemShut {NoStop}%
\bibitem [{\citenamefont {Solja\v{c}i{\'{c}}}\ \emph
  {et~al.}(2003)\citenamefont {Solja\v{c}i{\'{c}}}, \citenamefont {Luo},
  \citenamefont {Joannopoulos},\ and\ \citenamefont {Fan}}]{Soljacic2003}%
  \BibitemOpen
  \bibfield  {author} {\bibinfo {author} {\bibfnamefont {M.}~\bibnamefont
  {Solja\v{c}i{\'{c}}}}, \bibinfo {author} {\bibfnamefont {C.}~\bibnamefont
  {Luo}}, \bibinfo {author} {\bibfnamefont {J.~D.}\ \bibnamefont
  {Joannopoulos}}, \ and\ \bibinfo {author} {\bibfnamefont {S.}~\bibnamefont
  {Fan}},\ }\bibfield  {title} {\enquote {\bibinfo {title} {{Nonlinear photonic
  crystal microdevices for optical integration}},}\ }\href@noop {} {\bibfield
  {journal} {\bibinfo  {journal} {Optics Lett.}\ }\textbf {\bibinfo {volume}
  {28}},\ \bibinfo {pages} {637} (\bibinfo {year} {2003})}\BibitemShut
  {NoStop}%
\bibitem [{\citenamefont {Fan}\ \emph {et~al.}(2012)\citenamefont {Fan},
  \citenamefont {Wang}, \citenamefont {Varghese}, \citenamefont {Shen},
  \citenamefont {Niu}, \citenamefont {Xuan}, \citenamefont {Weiner},\ and\
  \citenamefont {Qi}}]{Fan2012}%
  \BibitemOpen
  \bibfield  {author} {\bibinfo {author} {\bibfnamefont {L.}~\bibnamefont
  {Fan}}, \bibinfo {author} {\bibfnamefont {J.}~\bibnamefont {Wang}}, \bibinfo
  {author} {\bibfnamefont {L.~T.}\ \bibnamefont {Varghese}}, \bibinfo {author}
  {\bibfnamefont {H.}~\bibnamefont {Shen}}, \bibinfo {author} {\bibfnamefont
  {B.}~\bibnamefont {Niu}}, \bibinfo {author} {\bibfnamefont {Y.}~\bibnamefont
  {Xuan}}, \bibinfo {author} {\bibfnamefont {A.~M.}\ \bibnamefont {Weiner}}, \
  and\ \bibinfo {author} {\bibfnamefont {M.}~\bibnamefont {Qi}},\ }\bibfield
  {title} {\enquote {\bibinfo {title} {{An All-Silicon Passive Optical
  Diode}},}\ }\href {\doibase 10.1126/science.1214383} {\bibfield  {journal}
  {\bibinfo  {journal} {Science}\ }\textbf {\bibinfo {volume} {335}},\ \bibinfo
  {pages} {447} (\bibinfo {year} {2012})}\BibitemShut {NoStop}%
\bibitem [{\citenamefont {Sounas}\ and\ \citenamefont
  {Al{\`{u}}}(2014)}]{Sounas2014}%
  \BibitemOpen
  \bibfield  {author} {\bibinfo {author} {\bibfnamefont {D.~L.}\ \bibnamefont
  {Sounas}}\ and\ \bibinfo {author} {\bibfnamefont {A.}~\bibnamefont
  {Al{\`{u}}}},\ }\bibfield  {title} {\enquote {\bibinfo {title}
  {{Angular-Momentum-Biased Nanorings to Realize Magnetic-Free Integrated
  Optical Isolation}},}\ }\href@noop {} {\bibfield  {journal} {\bibinfo
  {journal} {ACS Photonics}\ }\textbf {\bibinfo {volume} {1}},\ \bibinfo
  {pages} {198} (\bibinfo {year} {2014})}\BibitemShut {NoStop}%
\bibitem [{\citenamefont {Wu}\ \emph {et~al.}(2015)\citenamefont {Wu},
  \citenamefont {Chen}, \citenamefont {Ji}, \citenamefont {Huang},
  \citenamefont {Xia}, \citenamefont {Wu},\ and\ \citenamefont
  {Wang}}]{Wu2015}%
  \BibitemOpen
  \bibfield  {author} {\bibinfo {author} {\bibfnamefont {Z.}~\bibnamefont
  {Wu}}, \bibinfo {author} {\bibfnamefont {J.}~\bibnamefont {Chen}}, \bibinfo
  {author} {\bibfnamefont {M.}~\bibnamefont {Ji}}, \bibinfo {author}
  {\bibfnamefont {Q.}~\bibnamefont {Huang}}, \bibinfo {author} {\bibfnamefont
  {J.}~\bibnamefont {Xia}}, \bibinfo {author} {\bibfnamefont {Y.}~\bibnamefont
  {Wu}}, \ and\ \bibinfo {author} {\bibfnamefont {Y.}~\bibnamefont {Wang}},\
  }\bibfield  {title} {\enquote {\bibinfo {title} {{Optical nonreciprocal
  transmission in an asymmetric silicon photonic crystal structure}},}\
  }\href@noop {} {\bibfield  {journal} {\bibinfo  {journal} {Appl. Phys.
  Lett.}\ }\textbf {\bibinfo {volume} {107}},\ \bibinfo {pages} {221102}
  (\bibinfo {year} {2015})}\BibitemShut {NoStop}%
\bibitem [{\citenamefont {Guo}\ \emph {et~al.}(2016)\citenamefont {Guo},
  \citenamefont {Zou}, \citenamefont {Jung},\ and\ \citenamefont
  {Tang}}]{Guo2016}%
  \BibitemOpen
  \bibfield  {author} {\bibinfo {author} {\bibfnamefont {X.}~\bibnamefont
  {Guo}}, \bibinfo {author} {\bibfnamefont {C.-L.}\ \bibnamefont {Zou}},
  \bibinfo {author} {\bibfnamefont {H.}~\bibnamefont {Jung}}, \ and\ \bibinfo
  {author} {\bibfnamefont {H.~X.}\ \bibnamefont {Tang}},\ }\bibfield  {title}
  {\enquote {\bibinfo {title} {{On-Chip Strong Coupling and Efficient Frequency
  Conversion between Telecom and Visible Optical Modes}},}\ }\href {\doibase
  10.1103/PhysRevLett.117.123902} {\bibfield  {journal} {\bibinfo  {journal}
  {Phys. Rev. Lett.}\ }\textbf {\bibinfo {volume} {117}},\ \bibinfo {pages}
  {123902} (\bibinfo {year} {2016})},\ \Eprint
  {http://arxiv.org/abs/1511.08112} {1511.08112} \BibitemShut {NoStop}%
\bibitem [{\citenamefont {Hua}\ \emph {et~al.}(2016)\citenamefont {Hua},
  \citenamefont {Wen}, \citenamefont {Jiang}, \citenamefont {Hua},
  \citenamefont {Jiang},\ and\ \citenamefont {Xiao}}]{Hua2016}%
  \BibitemOpen
  \bibfield  {author} {\bibinfo {author} {\bibfnamefont {S.}~\bibnamefont
  {Hua}}, \bibinfo {author} {\bibfnamefont {J.}~\bibnamefont {Wen}}, \bibinfo
  {author} {\bibfnamefont {X.}~\bibnamefont {Jiang}}, \bibinfo {author}
  {\bibfnamefont {Q.}~\bibnamefont {Hua}}, \bibinfo {author} {\bibfnamefont
  {L.}~\bibnamefont {Jiang}}, \ and\ \bibinfo {author} {\bibfnamefont
  {M.}~\bibnamefont {Xiao}},\ }\bibfield  {title} {\enquote {\bibinfo {title}
  {{Demonstration of a chip-based optical isolator with parametric
  amplification}},}\ }\href@noop {} {\bibfield  {journal} {\bibinfo  {journal}
  {Nat. Commun.}\ }\textbf {\bibinfo {volume} {7}},\ \bibinfo {pages} {13657}
  (\bibinfo {year} {2016})}\BibitemShut {NoStop}%
\bibitem [{\citenamefont {Yu}\ and\ \citenamefont {Fan}(2009)}]{Yu2009}%
  \BibitemOpen
  \bibfield  {author} {\bibinfo {author} {\bibfnamefont {Z.}~\bibnamefont
  {Yu}}\ and\ \bibinfo {author} {\bibfnamefont {S.}~\bibnamefont {Fan}},\
  }\bibfield  {title} {\enquote {\bibinfo {title} {{Complete optical isolation
  created by indirect interband photonic transitions}},}\ }\href {\doibase
  10.1038/nphoton.2008.273} {\bibfield  {journal} {\bibinfo  {journal} {Nat.
  Photonics}\ }\textbf {\bibinfo {volume} {3}},\ \bibinfo {pages} {91}
  (\bibinfo {year} {2009})}\BibitemShut {NoStop}%
\bibitem [{\citenamefont {Lira}\ \emph {et~al.}(2012)\citenamefont {Lira},
  \citenamefont {Yu}, \citenamefont {Fan},\ and\ \citenamefont
  {Lipson}}]{Lira2012}%
  \BibitemOpen
  \bibfield  {author} {\bibinfo {author} {\bibfnamefont {H.}~\bibnamefont
  {Lira}}, \bibinfo {author} {\bibfnamefont {Z.}~\bibnamefont {Yu}}, \bibinfo
  {author} {\bibfnamefont {S.}~\bibnamefont {Fan}}, \ and\ \bibinfo {author}
  {\bibfnamefont {M.}~\bibnamefont {Lipson}},\ }\bibfield  {title} {\enquote
  {\bibinfo {title} {{Electrically driven nonreciprocity induced by interband
  photonic transition on a silicon chip}},}\ }\href@noop {} {\bibfield
  {journal} {\bibinfo  {journal} {Phys. Rev. Lett.}\ }\textbf {\bibinfo
  {volume} {109}},\ \bibinfo {pages} {033901} (\bibinfo {year}
  {2012})}\BibitemShut {NoStop}%
\bibitem [{\citenamefont {Estep}\ \emph {et~al.}(2014)\citenamefont {Estep},
  \citenamefont {Sounas}, \citenamefont {Soric},\ and\ \citenamefont
  {Al{\`{u}}}}]{Estep2014}%
  \BibitemOpen
  \bibfield  {author} {\bibinfo {author} {\bibfnamefont {N.~A.}\ \bibnamefont
  {Estep}}, \bibinfo {author} {\bibfnamefont {D.~L.}\ \bibnamefont {Sounas}},
  \bibinfo {author} {\bibfnamefont {J.}~\bibnamefont {Soric}}, \ and\ \bibinfo
  {author} {\bibfnamefont {A.}~\bibnamefont {Al{\`{u}}}},\ }\bibfield  {title}
  {\enquote {\bibinfo {title} {{Magnetic-free non-reciprocity and isolation
  based on parametrically modulated coupled-resonator loops}},}\ }\href
  {\doibase 10.1038/nphys3134} {\bibfield  {journal} {\bibinfo  {journal} {Nat.
  Phys.}\ }\textbf {\bibinfo {volume} {10}},\ \bibinfo {pages} {923} (\bibinfo
  {year} {2014})}\BibitemShut {NoStop}%
\bibitem [{\citenamefont {Yang}\ and\ \citenamefont {Li}(2016)}]{Yang2016}%
  \BibitemOpen
  \bibfield  {author} {\bibinfo {author} {\bibfnamefont {F.}~\bibnamefont
  {Yang}}\ and\ \bibinfo {author} {\bibfnamefont {Y.}~\bibnamefont {Li}},\
  }\bibfield  {title} {\enquote {\bibinfo {title} {{Nonreciprocal diffraction
  of light based on double-transition-assisted photonic Aharonov-Bohm
  effect}},}\ }\href@noop {} {\bibfield  {journal} {\bibinfo  {journal} {Phys.
  Rev. B}\ }\textbf {\bibinfo {volume} {94}},\ \bibinfo {pages} {165439}
  (\bibinfo {year} {2016})}\BibitemShut {NoStop}%
\bibitem [{\citenamefont {Hafezi}\ and\ \citenamefont
  {Rabl}(2012)}]{Hafezi2012}%
  \BibitemOpen
  \bibfield  {author} {\bibinfo {author} {\bibfnamefont {M.}~\bibnamefont
  {Hafezi}}\ and\ \bibinfo {author} {\bibfnamefont {P.}~\bibnamefont {Rabl}},\
  }\bibfield  {title} {\enquote {\bibinfo {title} {{Optomechanically induced
  non-reciprocity in microring resonators}},}\ }\href@noop {} {\bibfield
  {journal} {\bibinfo  {journal} {Opt. Express}\ }\textbf {\bibinfo {volume}
  {20}},\ \bibinfo {pages} {7672} (\bibinfo {year} {2012})}\BibitemShut
  {NoStop}%
\bibitem [{\citenamefont {Shen}\ \emph {et~al.}(2016)\citenamefont {Shen},
  \citenamefont {Zhang}, \citenamefont {Chen}, \citenamefont {Zou},
  \citenamefont {Xiao}, \citenamefont {Zou}, \citenamefont {Sun}, \citenamefont
  {Guo},\ and\ \citenamefont {Dong}}]{Shen2016}%
  \BibitemOpen
  \bibfield  {author} {\bibinfo {author} {\bibfnamefont {Z.}~\bibnamefont
  {Shen}}, \bibinfo {author} {\bibfnamefont {Y.-L.}\ \bibnamefont {Zhang}},
  \bibinfo {author} {\bibfnamefont {Y.}~\bibnamefont {Chen}}, \bibinfo {author}
  {\bibfnamefont {C.-L.}\ \bibnamefont {Zou}}, \bibinfo {author} {\bibfnamefont
  {Y.-F.}\ \bibnamefont {Xiao}}, \bibinfo {author} {\bibfnamefont {X.-B.}\
  \bibnamefont {Zou}}, \bibinfo {author} {\bibfnamefont {F.-W.}\ \bibnamefont
  {Sun}}, \bibinfo {author} {\bibfnamefont {G.-C.}\ \bibnamefont {Guo}}, \ and\
  \bibinfo {author} {\bibfnamefont {C.-H.}\ \bibnamefont {Dong}},\ }\bibfield
  {title} {\enquote {\bibinfo {title} {{Experimental realization of
  optomechanically induced non-reciprocity}},}\ }\href@noop {} {\bibfield
  {journal} {\bibinfo  {journal} {Nat. Photonics}\ }\textbf {\bibinfo {volume}
  {10}},\ \bibinfo {pages} {657} (\bibinfo {year} {2016})}\BibitemShut
  {NoStop}%
\bibitem [{\citenamefont {Ruesink}\ \emph {et~al.}(2016)\citenamefont
  {Ruesink}, \citenamefont {Miri}, \citenamefont {Al{\`u}},\ and\ \citenamefont
  {Verhagen}}]{Ruesink2016}%
  \BibitemOpen
  \bibfield  {author} {\bibinfo {author} {\bibfnamefont {F.}~\bibnamefont
  {Ruesink}}, \bibinfo {author} {\bibfnamefont {M.-A.}\ \bibnamefont {Miri}},
  \bibinfo {author} {\bibfnamefont {A.}~\bibnamefont {Al{\`u}}}, \ and\
  \bibinfo {author} {\bibfnamefont {E.}~\bibnamefont {Verhagen}},\ }\bibfield
  {title} {\enquote {\bibinfo {title} {{Nonreciprocity and magnetic-free
  isolation based on optomechanical interactions}},}\ }\href@noop {} {\bibfield
   {journal} {\bibinfo  {journal} {Nat. Commun.}\ }\textbf {\bibinfo {volume}
  {7}},\ \bibinfo {pages} {13662} (\bibinfo {year} {2016})}\BibitemShut
  {NoStop}%
\bibitem [{\citenamefont {Shen}\ \emph {et~al.}(2018)\citenamefont {Shen},
  \citenamefont {Zhang}, \citenamefont {Chen}, \citenamefont {Sun},
  \citenamefont {Zou}, \citenamefont {Guo}, \citenamefont {Zou},\ and\
  \citenamefont {Dong}}]{Shen2018}%
  \BibitemOpen
  \bibfield  {author} {\bibinfo {author} {\bibfnamefont {Z.}~\bibnamefont
  {Shen}}, \bibinfo {author} {\bibfnamefont {Y.-L.}\ \bibnamefont {Zhang}},
  \bibinfo {author} {\bibfnamefont {Y.}~\bibnamefont {Chen}}, \bibinfo {author}
  {\bibfnamefont {F.-W.}\ \bibnamefont {Sun}}, \bibinfo {author} {\bibfnamefont
  {X.-B.}\ \bibnamefont {Zou}}, \bibinfo {author} {\bibfnamefont {G.-C.}\
  \bibnamefont {Guo}}, \bibinfo {author} {\bibfnamefont {C.-L.}\ \bibnamefont
  {Zou}}, \ and\ \bibinfo {author} {\bibfnamefont {C.-H.}\ \bibnamefont
  {Dong}},\ }\bibfield  {title} {\enquote {\bibinfo {title} {{Reconfigurable
  optomechanical circulator and directional amplifier}},}\ }\href@noop {}
  {\bibfield  {journal} {\bibinfo  {journal} {Nat. Commun.}\ }\textbf {\bibinfo
  {volume} {9}},\ \bibinfo {pages} {1797} (\bibinfo {year} {2018})}\BibitemShut
  {NoStop}%
\bibitem [{\citenamefont {Ruesink}\ \emph {et~al.}(2018)\citenamefont
  {Ruesink}, \citenamefont {Mathew}, \citenamefont {Miri}, \citenamefont
  {Al{\`u}},\ and\ \citenamefont {Verhagen}}]{Ruesink2018}%
  \BibitemOpen
  \bibfield  {author} {\bibinfo {author} {\bibfnamefont {F.}~\bibnamefont
  {Ruesink}}, \bibinfo {author} {\bibfnamefont {J.~P.}\ \bibnamefont {Mathew}},
  \bibinfo {author} {\bibfnamefont {M.-A.}\ \bibnamefont {Miri}}, \bibinfo
  {author} {\bibfnamefont {A.}~\bibnamefont {Al{\`u}}}, \ and\ \bibinfo
  {author} {\bibfnamefont {E.}~\bibnamefont {Verhagen}},\ }\bibfield  {title}
  {\enquote {\bibinfo {title} {{Optical circulation in a multimode
  optomechanical resonator}},}\ }\href@noop {} {\bibfield  {journal} {\bibinfo
  {journal} {Nat. Commun.}\ }\textbf {\bibinfo {volume} {9}},\ \bibinfo {pages}
  {1798} (\bibinfo {year} {2018})}\BibitemShut {NoStop}%
\bibitem [{\citenamefont {Ranzani}\ and\ \citenamefont
  {Aumentado}(2015)}]{Ranzani2015}%
  \BibitemOpen
  \bibfield  {author} {\bibinfo {author} {\bibfnamefont {L.}~\bibnamefont
  {Ranzani}}\ and\ \bibinfo {author} {\bibfnamefont {J.}~\bibnamefont
  {Aumentado}},\ }\bibfield  {title} {\enquote {\bibinfo {title} {{Graph-based
  analysis of nonreciprocity in coupled-mode systems}},}\ }\href {\doibase
  10.1088/1367-2630/17/2/023024} {\bibfield  {journal} {\bibinfo  {journal}
  {New J. Phys.}\ }\textbf {\bibinfo {volume} {17}},\ \bibinfo {pages} {023024}
  (\bibinfo {year} {2015})}\BibitemShut {NoStop}%
\bibitem [{\citenamefont {Metelmann}\ and\ \citenamefont
  {Clerk}(2015)}]{Metelmann2015}%
  \BibitemOpen
  \bibfield  {author} {\bibinfo {author} {\bibfnamefont {A}~\bibnamefont
  {Metelmann}}\ and\ \bibinfo {author} {\bibfnamefont {A~A}\ \bibnamefont
  {Clerk}},\ }\bibfield  {title} {\enquote {\bibinfo {title} {{Nonreciprocal
  Photon Transmission and Amplification via Reservoir Engineering}},}\
  }\href@noop {} {\bibfield  {journal} {\bibinfo  {journal} {Phys. Rev. X}\
  }\textbf {\bibinfo {volume} {5}},\ \bibinfo {pages} {021025} (\bibinfo {year}
  {2015})}\BibitemShut {NoStop}%
\bibitem [{\citenamefont {Fang}\ \emph {et~al.}(2017)\citenamefont {Fang},
  \citenamefont {Luo}, \citenamefont {Metelmann}, \citenamefont {Matheny},
  \citenamefont {Marquardt}, \citenamefont {Clerk},\ and\ \citenamefont
  {Painter}}]{Fang2017}%
  \BibitemOpen
  \bibfield  {author} {\bibinfo {author} {\bibfnamefont {K.}~\bibnamefont
  {Fang}}, \bibinfo {author} {\bibfnamefont {J.}~\bibnamefont {Luo}}, \bibinfo
  {author} {\bibfnamefont {A.}~\bibnamefont {Metelmann}}, \bibinfo {author}
  {\bibfnamefont {M.~H.}\ \bibnamefont {Matheny}}, \bibinfo {author}
  {\bibfnamefont {F.}~\bibnamefont {Marquardt}}, \bibinfo {author}
  {\bibfnamefont {A.~A.}\ \bibnamefont {Clerk}}, \ and\ \bibinfo {author}
  {\bibfnamefont {O.}~\bibnamefont {Painter}},\ }\bibfield  {title} {\enquote
  {\bibinfo {title} {{Generalized non-reciprocity in an optomechanical circuit
  via synthetic magnetism and reservoir engineering}},}\ }\href@noop {}
  {\bibfield  {journal} {\bibinfo  {journal} {Nat. Phys.}\ }\textbf {\bibinfo
  {volume} {13}},\ \bibinfo {pages} {465} (\bibinfo {year} {2017})}\BibitemShut
  {NoStop}%
\bibitem [{\citenamefont {Li}\ \emph {et~al.}(2017)\citenamefont {Li},
  \citenamefont {Huang}, \citenamefont {Zhang},\ and\ \citenamefont
  {Tian}}]{Li2017}%
  \BibitemOpen
  \bibfield  {author} {\bibinfo {author} {\bibfnamefont {Y.}~\bibnamefont
  {Li}}, \bibinfo {author} {\bibfnamefont {Y.~Y.}\ \bibnamefont {Huang}},
  \bibinfo {author} {\bibfnamefont {X.~Z.}\ \bibnamefont {Zhang}}, \ and\
  \bibinfo {author} {\bibfnamefont {L.}~\bibnamefont {Tian}},\ }\bibfield
  {title} {\enquote {\bibinfo {title} {{Optical directional amplification in a
  three-mode optomechanical system}},}\ }\href@noop {} {\bibfield  {journal}
  {\bibinfo  {journal} {Opt. Express}\ }\textbf {\bibinfo {volume} {25}},\
  \bibinfo {pages} {18907} (\bibinfo {year} {2017})}\BibitemShut {NoStop}%
\bibitem [{\citenamefont {Tian}\ and\ \citenamefont {Li}(2017)}]{Tian2017}%
  \BibitemOpen
  \bibfield  {author} {\bibinfo {author} {\bibfnamefont {L.}~\bibnamefont
  {Tian}}\ and\ \bibinfo {author} {\bibfnamefont {Z.}~\bibnamefont {Li}},\
  }\bibfield  {title} {\enquote {\bibinfo {title} {{Nonreciprocal quantum-state
  conversion between microwave and optical photons}},}\ }\href@noop {}
  {\bibfield  {journal} {\bibinfo  {journal} {Phys. Rev. A}\ }\textbf {\bibinfo
  {volume} {96}},\ \bibinfo {pages} {013808} (\bibinfo {year}
  {2017})}\BibitemShut {NoStop}%
\bibitem [{\citenamefont {Jiang}\ \emph {et~al.}(2018)\citenamefont {Jiang},
  \citenamefont {Song},\ and\ \citenamefont {Li}}]{Cheng2018}%
  \BibitemOpen
  \bibfield  {author} {\bibinfo {author} {\bibfnamefont {C.}~\bibnamefont
  {Jiang}}, \bibinfo {author} {\bibfnamefont {L.~N.}\ \bibnamefont {Song}}, \
  and\ \bibinfo {author} {\bibfnamefont {Y.}~\bibnamefont {Li}},\ }\bibfield
  {title} {\enquote {\bibinfo {title} {{Directional amplifier in an
  optomechanical system with optical gain}},}\ }\href@noop {} {\bibfield
  {journal} {\bibinfo  {journal} {Phys. Rev. A}\ }\textbf {\bibinfo {volume}
  {97}},\ \bibinfo {pages} {053812} (\bibinfo {year} {2018})}\BibitemShut
  {NoStop}%
\bibitem [{\citenamefont {Li}\ \emph {et~al.}(2018)\citenamefont {Li},
  \citenamefont {Xiao}, \citenamefont {Li},\ and\ \citenamefont
  {Wang}}]{Li2018}%
  \BibitemOpen
  \bibfield  {author} {\bibinfo {author} {\bibfnamefont {G.}~\bibnamefont
  {Li}}, \bibinfo {author} {\bibfnamefont {X.}~\bibnamefont {Xiao}}, \bibinfo
  {author} {\bibfnamefont {Y.}~\bibnamefont {Li}}, \ and\ \bibinfo {author}
  {\bibfnamefont {X.}~\bibnamefont {Wang}},\ }\bibfield  {title} {\enquote
  {\bibinfo {title} {Tunable optical nonreciprocity and a phonon-photon router
  in an optomechanical system with coupled mechanical and optical modes},}\
  }\href@noop {} {\bibfield  {journal} {\bibinfo  {journal} {Phys. Rev. A}\
  }\textbf {\bibinfo {volume} {97}},\ \bibinfo {pages} {023801} (\bibinfo
  {year} {2018})}\BibitemShut {NoStop}%
\bibitem [{\citenamefont {Xu}\ \emph {et~al.}()\citenamefont {Xu},
  \citenamefont {Jiang}, \citenamefont {Clerk},\ and\ \citenamefont
  {Harris}}]{Xu2018}%
  \BibitemOpen
  \bibfield  {author} {\bibinfo {author} {\bibfnamefont {H.}~\bibnamefont
  {Xu}}, \bibinfo {author} {\bibfnamefont {L.}~\bibnamefont {Jiang}}, \bibinfo
  {author} {\bibfnamefont {A.~A.}\ \bibnamefont {Clerk}}, \ and\ \bibinfo
  {author} {\bibfnamefont {J.~G.~E.}\ \bibnamefont {Harris}},\ }\bibfield
  {title} {\enquote {\bibinfo {title} {{Nonreciprocal control and cooling of
  phonon modes in an optomechanical system}},}\ }\href@noop {} {\bibinfo
  {journal} {arXiv:1807.03484}\ }\BibitemShut {NoStop}%
\bibitem [{\citenamefont {Xu}\ and\ \citenamefont {Li}(2015)}]{XuXW2015}%
  \BibitemOpen
\bibfield  {journal} {  }\bibfield  {author} {\bibinfo {author} {\bibfnamefont
  {X.-W.}\ \bibnamefont {Xu}}\ and\ \bibinfo {author} {\bibfnamefont
  {Y.}~\bibnamefont {Li}},\ }\bibfield  {title} {\enquote {\bibinfo {title}
  {{Optical nonreciprocity and optomechanical circulator in three-mode
  optomechanical systems}},}\ }\href@noop {} {\bibfield  {journal} {\bibinfo
  {journal} {Phys. Rev. A}\ }\textbf {\bibinfo {volume} {91}},\ \bibinfo
  {pages} {053854} (\bibinfo {year} {2015})}\BibitemShut {NoStop}%
\bibitem [{\citenamefont {Xu}\ \emph {et~al.}(2016)\citenamefont {Xu},
  \citenamefont {Li}, \citenamefont {Chen},\ and\ \citenamefont
  {Liu}}]{XuXW2016}%
  \BibitemOpen
  \bibfield  {author} {\bibinfo {author} {\bibfnamefont {X.-W.}\ \bibnamefont
  {Xu}}, \bibinfo {author} {\bibfnamefont {Y.}~\bibnamefont {Li}}, \bibinfo
  {author} {\bibfnamefont {A.-X.}\ \bibnamefont {Chen}}, \ and\ \bibinfo
  {author} {\bibfnamefont {Y.-X.}\ \bibnamefont {Liu}},\ }\bibfield  {title}
  {\enquote {\bibinfo {title} {{Nonreciprocal conversion between microwave and
  optical photons in electro-optomechanical systems}},}\ }\href@noop {}
  {\bibfield  {journal} {\bibinfo  {journal} {Phys. Rev. A}\ }\textbf {\bibinfo
  {volume} {93}},\ \bibinfo {pages} {023827} (\bibinfo {year}
  {2016})}\BibitemShut {NoStop}%
\bibitem [{\citenamefont {Peterson}\ \emph {et~al.}(2017)\citenamefont
  {Peterson}, \citenamefont {Lecocq}, \citenamefont {Cicak}, \citenamefont
  {Simmonds}, \citenamefont {Aumentado},\ and\ \citenamefont
  {Teufel}}]{Peterson2017}%
  \BibitemOpen
  \bibfield  {author} {\bibinfo {author} {\bibfnamefont {G.~A.}\ \bibnamefont
  {Peterson}}, \bibinfo {author} {\bibfnamefont {F.}~\bibnamefont {Lecocq}},
  \bibinfo {author} {\bibfnamefont {K.}~\bibnamefont {Cicak}}, \bibinfo
  {author} {\bibfnamefont {R.~W.}\ \bibnamefont {Simmonds}}, \bibinfo {author}
  {\bibfnamefont {J.}~\bibnamefont {Aumentado}}, \ and\ \bibinfo {author}
  {\bibfnamefont {J.~D.}\ \bibnamefont {Teufel}},\ }\bibfield  {title}
  {\enquote {\bibinfo {title} {Demonstration of efficient nonreciprocity in a
  microwave optomechanical circuit},}\ }\href@noop {} {\bibfield  {journal}
  {\bibinfo  {journal} {Phys. Rev. X}\ }\textbf {\bibinfo {volume} {7}},\
  \bibinfo {pages} {031001} (\bibinfo {year} {2017})}\BibitemShut {NoStop}%
\bibitem [{\citenamefont {Bernier}\ \emph {et~al.}(2017)\citenamefont
  {Bernier}, \citenamefont {T\'oth}, \citenamefont {Koottandavida},
  \citenamefont {Ioannou}, \citenamefont {Malz}, \citenamefont {Nunnenkamp},
  \citenamefont {Feofanov},\ and\ \citenamefont {Kippenberg}}]{Bernier2017}%
  \BibitemOpen
  \bibfield  {author} {\bibinfo {author} {\bibfnamefont {N.~R.}\ \bibnamefont
  {Bernier}}, \bibinfo {author} {\bibfnamefont {L.~D.}\ \bibnamefont {T\'oth}},
  \bibinfo {author} {\bibfnamefont {A.}~\bibnamefont {Koottandavida}}, \bibinfo
  {author} {\bibfnamefont {M.~A.}\ \bibnamefont {Ioannou}}, \bibinfo {author}
  {\bibfnamefont {D.}~\bibnamefont {Malz}}, \bibinfo {author} {\bibfnamefont
  {A.}~\bibnamefont {Nunnenkamp}}, \bibinfo {author} {\bibfnamefont {A.~K.}\
  \bibnamefont {Feofanov}}, \ and\ \bibinfo {author} {\bibfnamefont {T.~J.}\
  \bibnamefont {Kippenberg}},\ }\bibfield  {title} {\enquote {\bibinfo {title}
  {{Nonreciprocal reconfigurable microwave optomechanical circuit}},}\
  }\href@noop {} {\bibfield  {journal} {\bibinfo  {journal} {Nat. Commun.}\
  }\textbf {\bibinfo {volume} {8}},\ \bibinfo {pages} {604} (\bibinfo {year}
  {2017})}\BibitemShut {NoStop}%
\bibitem [{\citenamefont {Barzanjeh}\ \emph {et~al.}(2017)\citenamefont
  {Barzanjeh}, \citenamefont {Wulf}, \citenamefont {Peruzzo}, \citenamefont
  {Kalaee}, \citenamefont {Dieterle}, \citenamefont {Painter},\ and\
  \citenamefont {Fink}}]{Barzanjeh2017}%
  \BibitemOpen
  \bibfield  {author} {\bibinfo {author} {\bibfnamefont {S.}~\bibnamefont
  {Barzanjeh}}, \bibinfo {author} {\bibfnamefont {M.}~\bibnamefont {Wulf}},
  \bibinfo {author} {\bibfnamefont {M.}~\bibnamefont {Peruzzo}}, \bibinfo
  {author} {\bibfnamefont {M.}~\bibnamefont {Kalaee}}, \bibinfo {author}
  {\bibfnamefont {P.~B.}\ \bibnamefont {Dieterle}}, \bibinfo {author}
  {\bibfnamefont {O.}~\bibnamefont {Painter}}, \ and\ \bibinfo {author}
  {\bibfnamefont {J.~M.}\ \bibnamefont {Fink}},\ }\bibfield  {title} {\enquote
  {\bibinfo {title} {{Mechanical on-chip microwave circulator}},}\ }\href@noop
  {} {\bibfield  {journal} {\bibinfo  {journal} {Nat. Commun.}\ }\textbf
  {\bibinfo {volume} {8}},\ \bibinfo {pages} {953} (\bibinfo {year}
  {2017})}\BibitemShut {NoStop}%
\bibitem [{\citenamefont {Massel}\ \emph {et~al.}(2011)\citenamefont {Massel},
  \citenamefont {Heikkil{\"a}}, \citenamefont {Pirkkalainen}, \citenamefont
  {Cho}, \citenamefont {Saloniemi}, \citenamefont {Hakonen},\ and\
  \citenamefont {Sillanp{\"a}{\"a}}}]{MechAmpPaper}%
  \BibitemOpen
  \bibfield  {author} {\bibinfo {author} {\bibfnamefont {F.}~\bibnamefont
  {Massel}}, \bibinfo {author} {\bibfnamefont {T.~T.}\ \bibnamefont
  {Heikkil{\"a}}}, \bibinfo {author} {\bibfnamefont {J.-M.}\ \bibnamefont
  {Pirkkalainen}}, \bibinfo {author} {\bibfnamefont {S.~U.}\ \bibnamefont
  {Cho}}, \bibinfo {author} {\bibfnamefont {H.}~\bibnamefont {Saloniemi}},
  \bibinfo {author} {\bibfnamefont {P.~J.}\ \bibnamefont {Hakonen}}, \ and\
  \bibinfo {author} {\bibfnamefont {M.~A.}\ \bibnamefont {Sillanp{\"a}{\"a}}},\
  }\bibfield  {title} {\enquote {\bibinfo {title} {Microwave amplification with
  nanomechanical resonators},}\ }\href@noop {} {\bibfield  {journal} {\bibinfo
  {journal} {Nature}\ }\textbf {\bibinfo {volume} {480}},\ \bibinfo {pages}
  {351--354} (\bibinfo {year} {2011})}\BibitemShut {NoStop}%
\bibitem [{\citenamefont {Ockeloen-Korppi}\ \emph {et~al.}(2016)\citenamefont
  {Ockeloen-Korppi}, \citenamefont {Damsk\"agg}, \citenamefont {Pirkkalainen},
  \citenamefont {Heikkil\"a}, \citenamefont {Massel},\ and\ \citenamefont
  {Sillanp\"a\"a}}]{CasparAmp}%
  \BibitemOpen
  \bibfield  {author} {\bibinfo {author} {\bibfnamefont {C.~F.}\ \bibnamefont
  {Ockeloen-Korppi}}, \bibinfo {author} {\bibfnamefont {E.}~\bibnamefont
  {Damsk\"agg}}, \bibinfo {author} {\bibfnamefont {J.-M.}\ \bibnamefont
  {Pirkkalainen}}, \bibinfo {author} {\bibfnamefont {T.~T.}\ \bibnamefont
  {Heikkil\"a}}, \bibinfo {author} {\bibfnamefont {F.}~\bibnamefont {Massel}},
  \ and\ \bibinfo {author} {\bibfnamefont {M.~A.}\ \bibnamefont
  {Sillanp\"a\"a}},\ }\bibfield  {title} {\enquote {\bibinfo {title} {Low-noise
  amplification and frequency conversion with a multiport microwave
  optomechanical device},}\ }\href@noop {} {\bibfield  {journal} {\bibinfo
  {journal} {Phys. Rev. X}\ }\textbf {\bibinfo {volume} {6}},\ \bibinfo {pages}
  {041024} (\bibinfo {year} {2016})}\BibitemShut {NoStop}%
\bibitem [{\citenamefont {T{\'o}th}\ \emph {et~al.}(2017)\citenamefont
  {T{\'o}th}, \citenamefont {Bernier}, \citenamefont {Nunnenkamp},
  \citenamefont {Feofanov},\ and\ \citenamefont {Kippenberg}}]{KippenbergAmp}%
  \BibitemOpen
  \bibfield  {author} {\bibinfo {author} {\bibfnamefont {L.~D.}\ \bibnamefont
  {T{\'o}th}}, \bibinfo {author} {\bibfnamefont {N.~R.}\ \bibnamefont
  {Bernier}}, \bibinfo {author} {\bibfnamefont {A.}~\bibnamefont {Nunnenkamp}},
  \bibinfo {author} {\bibfnamefont {A.~K.}\ \bibnamefont {Feofanov}}, \ and\
  \bibinfo {author} {\bibfnamefont {T.~J.}\ \bibnamefont {Kippenberg}},\
  }\bibfield  {title} {\enquote {\bibinfo {title} {{A dissipative quantum
  reservoir for microwave light using a mechanical oscillator}},}\ }\href@noop
  {} {\bibfield  {journal} {\bibinfo  {journal} {Nat. Phys.}\ }\textbf
  {\bibinfo {volume} {13}},\ \bibinfo {pages} {787} (\bibinfo {year}
  {2017})}\BibitemShut {NoStop}%
\bibitem [{\citenamefont {Ockeloen-Korppi}\ \emph
  {et~al.}(2017{\natexlab{a}})\citenamefont {Ockeloen-Korppi}, \citenamefont
  {Damsk\"agg}, \citenamefont {Pirkkalainen}, \citenamefont {Heikkil\"a},
  \citenamefont {Massel},\ and\ \citenamefont {Sillanp\"a\"a}}]{SqueezeAmp}%
  \BibitemOpen
  \bibfield  {author} {\bibinfo {author} {\bibfnamefont {C.~F.}\ \bibnamefont
  {Ockeloen-Korppi}}, \bibinfo {author} {\bibfnamefont {E.}~\bibnamefont
  {Damsk\"agg}}, \bibinfo {author} {\bibfnamefont {J.-M.}\ \bibnamefont
  {Pirkkalainen}}, \bibinfo {author} {\bibfnamefont {T.~T.}\ \bibnamefont
  {Heikkil\"a}}, \bibinfo {author} {\bibfnamefont {F.}~\bibnamefont {Massel}},
  \ and\ \bibinfo {author} {\bibfnamefont {M.~A.}\ \bibnamefont
  {Sillanp\"a\"a}},\ }\bibfield  {title} {\enquote {\bibinfo {title} {Noiseless
  quantum measurement and squeezing of microwave fields utilizing mechanical
  vibrations},}\ }\href@noop {} {\bibfield  {journal} {\bibinfo  {journal}
  {Phys. Rev. Lett.}\ }\textbf {\bibinfo {volume} {118}},\ \bibinfo {pages}
  {103601} (\bibinfo {year} {2017}{\natexlab{a}})}\BibitemShut {NoStop}%
\bibitem [{\citenamefont {Malz}\ \emph {et~al.}(2018)\citenamefont {Malz},
  \citenamefont {T\'oth}, \citenamefont {Bernier}, \citenamefont {Feofanov},
  \citenamefont {Kippenberg},\ and\ \citenamefont {Nunnenkamp}}]{Malz2018}%
  \BibitemOpen
  \bibfield  {author} {\bibinfo {author} {\bibfnamefont {D.}~\bibnamefont
  {Malz}}, \bibinfo {author} {\bibfnamefont {L.~D.}\ \bibnamefont {T\'oth}},
  \bibinfo {author} {\bibfnamefont {N.~R.}\ \bibnamefont {Bernier}}, \bibinfo
  {author} {\bibfnamefont {A.~K.}\ \bibnamefont {Feofanov}}, \bibinfo {author}
  {\bibfnamefont {T.~J.}\ \bibnamefont {Kippenberg}}, \ and\ \bibinfo {author}
  {\bibfnamefont {A.}~\bibnamefont {Nunnenkamp}},\ }\bibfield  {title}
  {\enquote {\bibinfo {title} {{Quantum-Limited Directional Amplifiers with
  Optomechanics}},}\ }\href@noop {} {\bibfield  {journal} {\bibinfo  {journal}
  {Phys. Rev. Lett.}\ }\textbf {\bibinfo {volume} {120}},\ \bibinfo {pages}
  {023601} (\bibinfo {year} {2018})}\BibitemShut {NoStop}%
\bibitem [{\citenamefont {Caves}(1982)}]{caves82}%
  \BibitemOpen
  \bibfield  {author} {\bibinfo {author} {\bibfnamefont {C.M.}\ \bibnamefont
  {Caves}},\ }\bibfield  {title} {\enquote {\bibinfo {title} {Quantum limits on
  noise in linear amplifiers},}\ }\href@noop {} {\bibfield  {journal} {\bibinfo
   {journal} {Phys. Rev. D}\ }\textbf {\bibinfo {volume} {26}},\ \bibinfo
  {pages} {1817} (\bibinfo {year} {1982})}\BibitemShut {NoStop}%
\bibitem [{\citenamefont {Ockeloen-Korppi}\ \emph
  {et~al.}(2017{\natexlab{b}})\citenamefont {Ockeloen-Korppi}, \citenamefont
  {Heikkil\"a}, \citenamefont {Sillanp\"a\"a},\ and\ \citenamefont
  {Massel}}]{SqAmpTheory}%
  \BibitemOpen
  \bibfield  {author} {\bibinfo {author} {\bibfnamefont {C.~F.}\ \bibnamefont
  {Ockeloen-Korppi}}, \bibinfo {author} {\bibfnamefont {T.~T.}\ \bibnamefont
  {Heikkil\"a}}, \bibinfo {author} {\bibfnamefont {M.~A.}\ \bibnamefont
  {Sillanp\"a\"a}}, \ and\ \bibinfo {author} {\bibfnamefont {F.}~\bibnamefont
  {Massel}},\ }\bibfield  {title} {\enquote {\bibinfo {title} {{Theory of
  phase-mixing amplification in an optomechanical system}},}\ }\href@noop {}
  {\bibfield  {journal} {\bibinfo  {journal} {Quantum Sci. Technol.}\ }\textbf
  {\bibinfo {volume} {2}},\ \bibinfo {pages} {035002} (\bibinfo {year}
  {2017}{\natexlab{b}})}\BibitemShut {NoStop}%
\bibitem [{\citenamefont {Teufel}\ \emph {et~al.}(2011)\citenamefont {Teufel},
  \citenamefont {Donner}, \citenamefont {Li}, \citenamefont {Harlow},
  \citenamefont {Allman}, \citenamefont {Cicak}, \citenamefont {Sirois},
  \citenamefont {Whittaker}, \citenamefont {Lehnert},\ and\ \citenamefont
  {Simmonds}}]{Teufel2011}%
  \BibitemOpen
  \bibfield  {author} {\bibinfo {author} {\bibfnamefont {J.~D.}\ \bibnamefont
  {Teufel}}, \bibinfo {author} {\bibfnamefont {T.}~\bibnamefont {Donner}},
  \bibinfo {author} {\bibfnamefont {D.}~\bibnamefont {Li}}, \bibinfo {author}
  {\bibfnamefont {J.~W.}\ \bibnamefont {Harlow}}, \bibinfo {author}
  {\bibfnamefont {M.~S.}\ \bibnamefont {Allman}}, \bibinfo {author}
  {\bibfnamefont {K.}~\bibnamefont {Cicak}}, \bibinfo {author} {\bibfnamefont
  {A.~J.}\ \bibnamefont {Sirois}}, \bibinfo {author} {\bibfnamefont {J.~D.}\
  \bibnamefont {Whittaker}}, \bibinfo {author} {\bibfnamefont {K.~W.}\
  \bibnamefont {Lehnert}}, \ and\ \bibinfo {author} {\bibfnamefont {R.~W.}\
  \bibnamefont {Simmonds}},\ }\bibfield  {title} {\enquote {\bibinfo {title}
  {{Sideband cooling of micromechanical motion to the quantum ground state}},}\
  }\href@noop {} {\bibfield  {journal} {\bibinfo  {journal} {Nature}\ }\textbf
  {\bibinfo {volume} {475}},\ \bibinfo {pages} {359} (\bibinfo {year}
  {2011})}\BibitemShut {NoStop}%
\end{thebibliography}

%

\end{document}